\documentclass[secnumarabic,aps,pra,amsfonts,twocolumn,%
showkeys,
]{revtex4}

\usepackage{bm,longtable,amsfonts}
\usepackage{graphicx}
\newcommand{\gl}[1]{(\ref{#1})}

\begin{document}

\title{Antiferromagnetic, Neutral, and Superconducting Band in
  La$_2$CuO$_4$} 
\author{Ekkehard Kr\"uger}
\email{krueger@mf.mpg.de}
\affiliation{Max-Planck-Institut f\"ur Metallforschung, D-70506 Stuttgart,
  Germany}
%
\date{\today}
\begin{abstract}
  The symmetry of the Bloch functions in the conduction band of tetragonal
  and orthorhombic La$_2$CuO$_4$ is examined for the existence of
  symmetry-adapted and optimally localizable (usual or spin-dependent)
  Wannier functions. It turns out that such Wannier functions do not exist
  in the tetragonal phase. In the orthorhombic phase, on the other hand, the
  Bloch functions can be unitarily transformed in three different ways into
  optimally localizable Wannier functions: they can be chosen to be adapted
  to each of the three phases observed in the pure or doped material, that
  is, to the antiferromagnetic phase, to the superconducting phase or to the
  phase evincing neither magnetism nor superconductivity. This
  group-theoretical result is proposed to be interpreted within a
  nonadiabatic extension of the Heisenberg model. Within this model,
  atomiclike states represented by these Wannier functions are responsible
  for the stability of each of the three phases.  However, all the three
  atomiclike states cannot exist in the tetragonal phase, but are stabilized
  by the orthorhombic distortion of the crystal. A simple model is proposed
  which might explain the physical properties of La$_{2-x}$Sr$_x$CuO$_4$ as
  a function of the Sr concentration $x$.
\end{abstract}
\maketitle

\section{Introduction}
\subsection{La$_{2-x}$Sr$_x$CuO$_4$}
The physical properties of pure and doped La$_{2-x}$Sr$_x$CuO$_4$ have been
studied in a broad range of Sr concentrations $x$ \cite{keimer,cava}. Pure
La$_{2}$CuO$_4$ is antiferromagnetic below about 300 K, but this magnetic
long-range order disappears at a Sr concentration of $x \approx 0.015$.
Between $x \approx 0.015$ and $x \approx 0.05$ the samples evince neither an
antiferromagnetic or a superconducting state. This border region has
features characteristic of spin-glass systems \cite{sternlieb}. Finally,
above $x \approx 0.05$ the material becomes a high-$T_c$ superconductor.

\begin{figure}[!]
\includegraphics[width=.3\textwidth,angle=0]{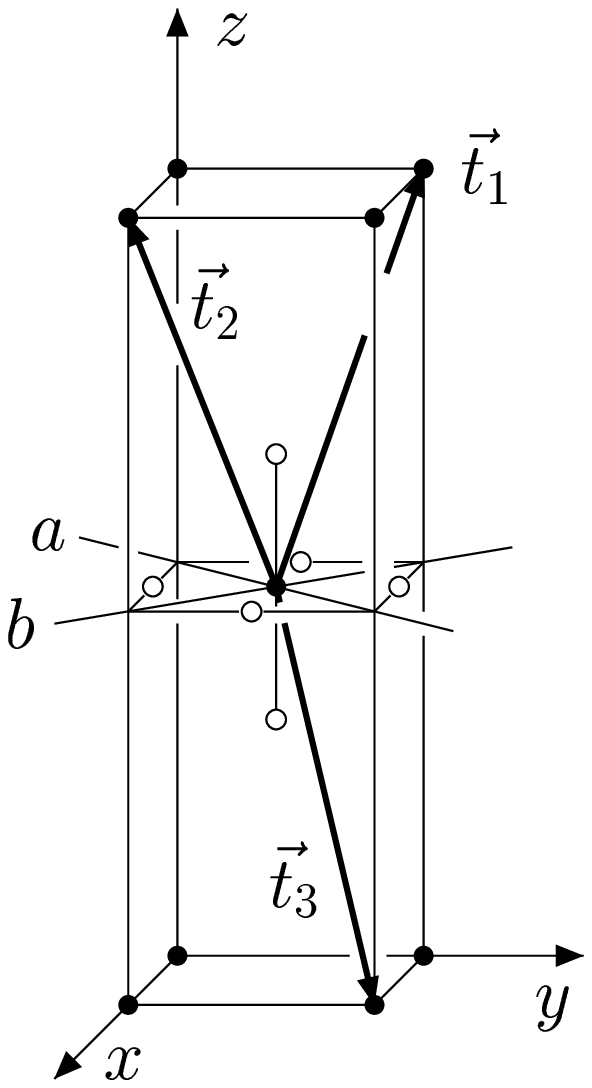}%
\caption{
  Structure of tetragonal La$_2$CuO$_4$, showing the basic translations of
  the Bravais lattice $\Gamma^v_q$. The space group is $I4/mmm$. Open
  circles are O, solid are Cu. La is not shown.
\label{fig:strukttetra}
}
\end{figure}

\begin{figure}[!]
\includegraphics[width=.28\textwidth,angle=0]{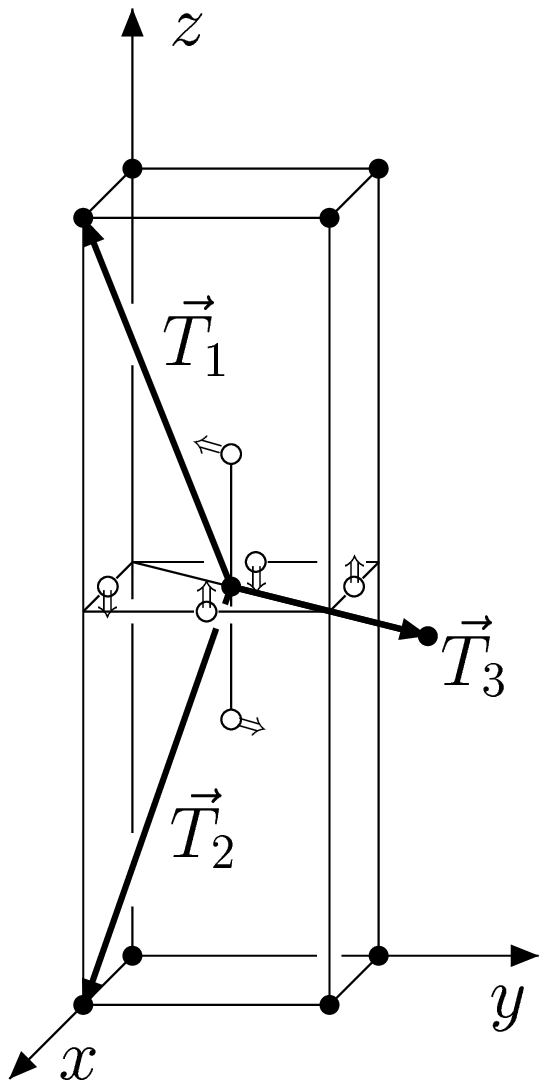}%
\caption{
  Structure of orthorhombic La$_2$CuO$_4$ with the space group $Cmca$,
  showing the basic translations of the Bravais lattice $\Gamma^b_o$. Open
  circles are O, solid are Cu. La is not shown. The arrows $\Uparrow{}$
  indicate the distortions of the O atoms. 
\label{fig:struktortho}
}
\end{figure}

Furthermore, La$_2$CuO$_4$ undergoes a very interesting structural phase
transition at about 520 K from a high-temperature tetragonal phase to a
low-temperature orthorhombic phase \cite{keimer}. The tetragonal structure
with the space group $I4/mmm$ is depicted in Fig.~\ref{fig:strukttetra}. The
major change in the structure in the tetragonal to orthorhombic transition
involves a ``buckling of the copper-oxygen planes'' \cite{cava}: in the
orthorhombic phase with the space group $Cmca$ the oxygen atoms are slightly
shifted from their tetragonal positions in such a way that the translation
$\vec t_3$ in Fig.~\ref{fig:strukttetra} is doubled \cite{cava,keimer}, see
Fig.~\ref{fig:struktortho}. The orthorhombic distortion is so small that the
orthorhombic and tetragonal structures are not distinguishable on the scale
of the figures \cite{cava}.

The buckling of the copper-oxygen planes has been established also in doped
La$_{2-x}$Sr$_x$CuO$_4$ for a broad range of Sr concentrations $x$. At zero
temperature, is exists up to $x \approx 0.2$ and disappears for higher Sr
concentrations \cite{keimer}.


\subsection{Magnetic and superconducting bands}

An energy band is called ``magnetic band with the magnetic group $M$'' if
the Bloch functions of this band can be unitarily transformed into Wannier
functions which are
\begin{itemize}
\item centered on the atomic positions;
\item symmetry-adapted to $M$; and
\item optimally localizable.
\end{itemize}
``Optimally localizable'' means that the Wannier functions are as well
localizable as possible.  Observations on the band structures of
Fe \cite{ef} and Cr \cite{ea} suggest that in both Fe and Cr a narrow,
roughly half-filled magnetic band is responsible for the stability of the
ferromagnetic and the (antiferromagnetic) spin-density wave state,
respectively, in these materials. This observation can be interpreted
within a nonadiabatic extension of the Heisenberg model, the ``nonadiabatic
Heisenberg model'' (NHM) \cite{enhm}. 

An energy band is called ``superconducting band'' if the Bloch functions of
this band can be unitarily transformed into {\em spin-dependent} Wannier
functions which are
\begin{itemize}
\item centered on the atomic positions;
\item symmetry-adapted to the space group of the considered material; and
\item optimally localizable. 
\end{itemize}
Observations on the band structures of a great number of elemental metals
suggest that narrow, roughly half-filled superconducting bands are
responsible for the stability of superconducting states. Also this
observation can be interpreted within the NHM \cite{josn,josm,josi}.

Optimally localizable and symmetry-adapted Wannier functions are defined in
Appendix \ref{sec:baender}.  The small irreducible representations of all
the magnetic or superconducting energy bands in the space groups of
La$_2$CuO$_4$ are listed in Tables~\ref{tab:wftetra} -- \ref{tab:afband} at
the end of this paper. The way to determine these symmetry labels is
also summarized in Appendix \ref{sec:baender}.

\subsection{Nonadiabatic Heisenberg model}

Within the NHM, the electrons in narrow, half-filled bands may lower their
Coulomb energy by occupying an atomiclike state as defined by Mott
\cite{mott} and Hubbard \cite{hubbard}: the electrons occupy as long as
possible localized states and perform their band motion by hopping from one
atom to another. The Hamiltonian $H^n$ within the NHM is distinguished from
any adiabatic Hamiltonian by a particular feature: the symmetry of $H^n$
depends on the symmetry of the localized functions representing the
localized states.  As a consequence, within this model the localized states
{\em must} be represented by optimally localizable Wannier functions which
are adapted to the symmetry of the considered phase.  These Wannier
functions may be usual (spin-independent) or spin-dependent Wannier
functions.

Spin-independent Wannier functions are related to magnetic bands, and, for
the first time in this paper, to a ``neutral'' band. In an atomiclike state
represented by {\em spin-dependent} Wannier functions, the conservation of
crystal spin requires a strong coupling between the electron spins and the
crystal spins of suitable boson excitations. Below a transition temperature,
this spin-boson interaction leads to stable Cooper pairs. 

The mechanism of Cooper pair formation within such an atomiclike system
resembles the familiar BCS mechanism \cite{bcs}.  However, the Cooper pairs
are stabilized in a new way by {\em constraining forces}.  Though these
constraining forces neither are considered within the original BCS theory
nor within the modern theory of superconductivity, it cannot be excluded
that they must operate in superconducting states because electrons forming
Cooper pairs possess only half of the degrees of freedom of free electrons.

The above mentioned ``suitable bosons'' are boson excitations bearing the
crystal-spin angular momentum $1\cdot\hbar$ and being sufficiently stable to
carry it through the crystal. The order of magnitude of the superconducting
transition temperature is determined by the excitation energy of those
stable crystal-spin-1 bosons of the crystal which have the lowest excitation
energy \cite{ehtc}.

All the Wannier functions considered in this paper are centered at the Cu
atoms because La$_{2}$CuO$_4$ has a single conduction band \cite{enhm}.

\section{Conduction band of tetragonal
  L\lowercase{a}$_2$C\lowercase{u}O$_4$} 
\begin{figure*}[!]
\includegraphics[width=.9\textwidth,angle=0]{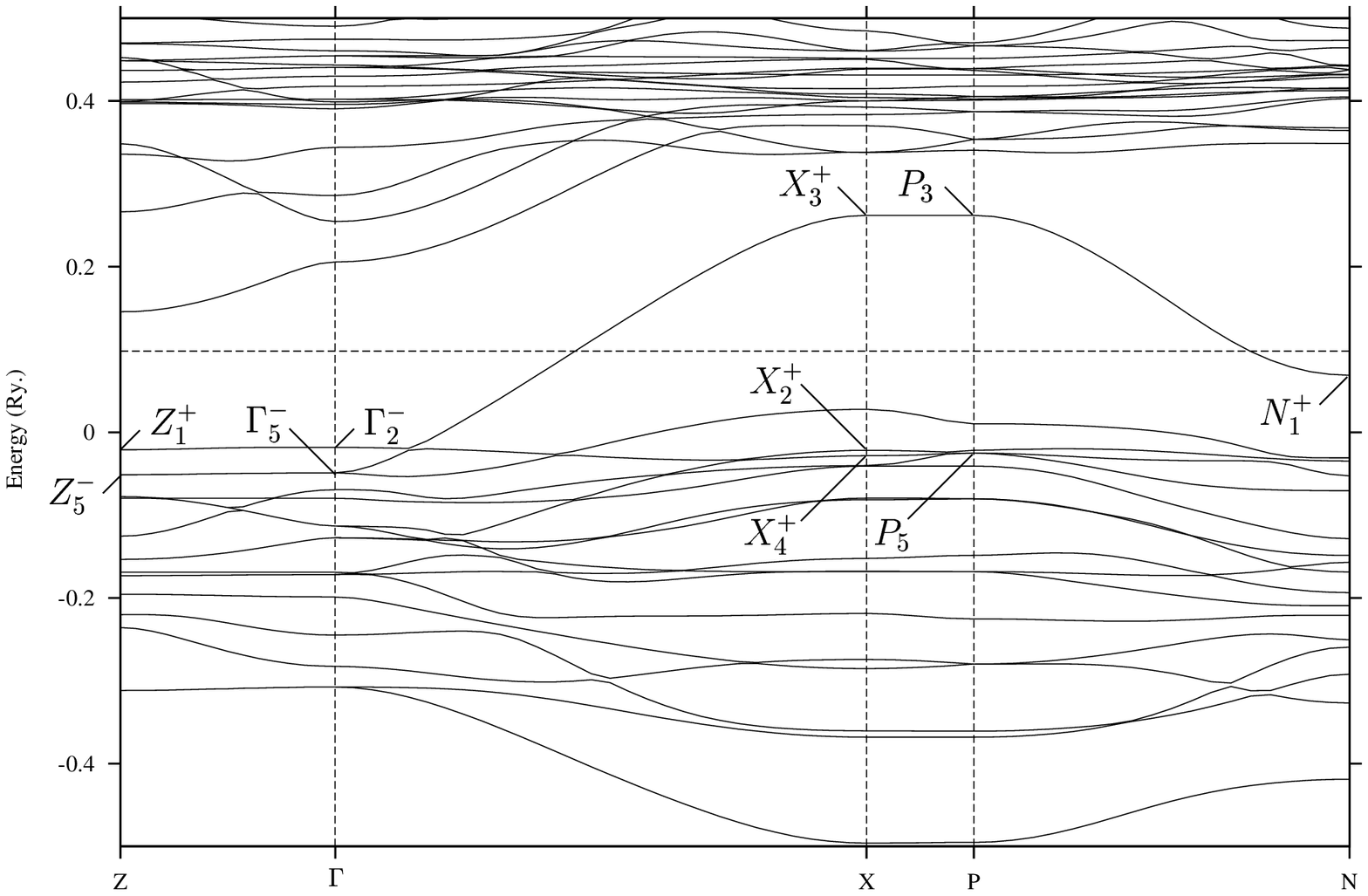}%
\caption{
  Band structure of tetragonal La$_2$CuO$_4$ as calculated by Ove Jepsen
  \cite{jepsen}, with symmetry labels determined by the author. The band
  characterized by $Z^+_1$, $\Gamma^-_2$ (or $Z^-_5$, $\Gamma^-_5$), $X^+_3
  + X^+_2$, $P_3 + P_5$, and $N^+_1$ forms a natural narrow conduction band.
  However, the electrons of this nearly half-filled band cannot occupy the
  (energetically favorable) atomiclike state because suitable
  symmetry-adapted and optimally localizable Wannier functions do not exist
  in the tetragonal phase. However, when the bands are folded into the
  Brillouin zone of the orthorhombic phase (see
  Fig.~\ref{fig:bandstrortho}), the points $X$ and $\Gamma$ 
  become equivalent. In the new Brillouin zone we may replace the $Z^+_1$
  and $\Gamma^-_2$ Bloch functions by the $X^+_2$ and $X^+_4$ functions of
  nearly the same energy.  With these new functions, the construction of
  symmetry-adapted Wannier becomes possible.
\label{fig:bandstrtetra}
}
\end{figure*}

Tetragonal La$_{2}$CuO$_4$ possesses a pronounced conduction band
characterized by the small representations
\begin{equation}
Z^+_1, \Gamma^-_2, X^+_3, P_3,\text{ and } N^+_1, 
\label{edtetra}
\end{equation}
see Fig.~\ref{fig:bandstrtetra}.

\subsection{Exclusion of a neutral atomiclike state with usual Wannier
  functions} Band \gl{edtetra} is not listed in Table \ref{tab:wftetra}.
Hence, in the tetragonal phase of La$_2$CuO$_4$ the electrons of
this band cannot lower their Coulomb energy by occupying an atomiclike state
because suitable symmetry-adapted and optimally localizable Wannier
functions do not exist.


\subsection{Exclusion of a superconducting band}
Taking into account the electron spin, the conduction band \gl{edtetra} is
characterized by the double-valued small representations
\begin{equation}
Z^+_6, \Gamma^-_6, X^+_5, P_7,\text{ and } N^+_3+N^+_4, 
\label{zdtetra}
\end{equation}
see Table \ref{tab:edzdtetra}. This band is not a superconducting band
because it is not listed in Table \ref{tab:slbandtetra}. Hence, the
electrons of this band cannot lower their Coulomb energy by occupying an
atomiclike state even if the Wannier functions representing the localized
states are allowed to be spin-dependent. Thus, in the framework of the NHM
we cannot expect stable Cooper pairs in the tetragonal phase of
La$_2$CuO$_4$.

\section{Exclusion of an antiferromagnetic band with the space group of
  antiferromagnetic chromium}  

Folding the conduction band \gl{edtetra} into the Brillouin zone of
antiferromagnetic Cr (as given, e.g., in Fig. 3 of Ref.\ \cite{eabf}), the
band becomes characterized by the small representations
\begin{equation}
\Gamma^+_2+\Gamma^-_2, M_{20}, A_{10},\text{ and } R_5, 
\label{edafcr}
\end{equation}
see Table \ref{tab:faltentetracraf}. This band is not listed in Table
\ref{tab:afbandcr}. Hence, optimally localizable Wannier functions being
adapted to $P4/mnc$ do not exist. The electrons cannot lower their Coulomb
energy by occupying an atomiclike state represented by such
antiferromagnetic Wannier functions. Hence, a spin structure with the space
group $P4/mnc$ (that means, with the spins lying parallel to the $z$ axis in
Fig.~\ref{fig:strukttetra}) is not predicted within the NHM.

\section{Conduction band of orthorhombic
  L\lowercase{a}$_2$C\lowercase{u}O$_4$} 
In the tetragonal phase of La$_2$CuO$_4$, it is the symmetry of the Bloch
functions at point $\Gamma$ which does not allow the construction of
symmetry-adapted and optimally localizable (spin-dependent) Wannier
functions: the Bloch functions at $\Gamma$ are negative with respect to the
inversion, but at the other points of symmetry they are positive. 


\begin{figure}[!]
   \includegraphics[width=.3\textwidth]{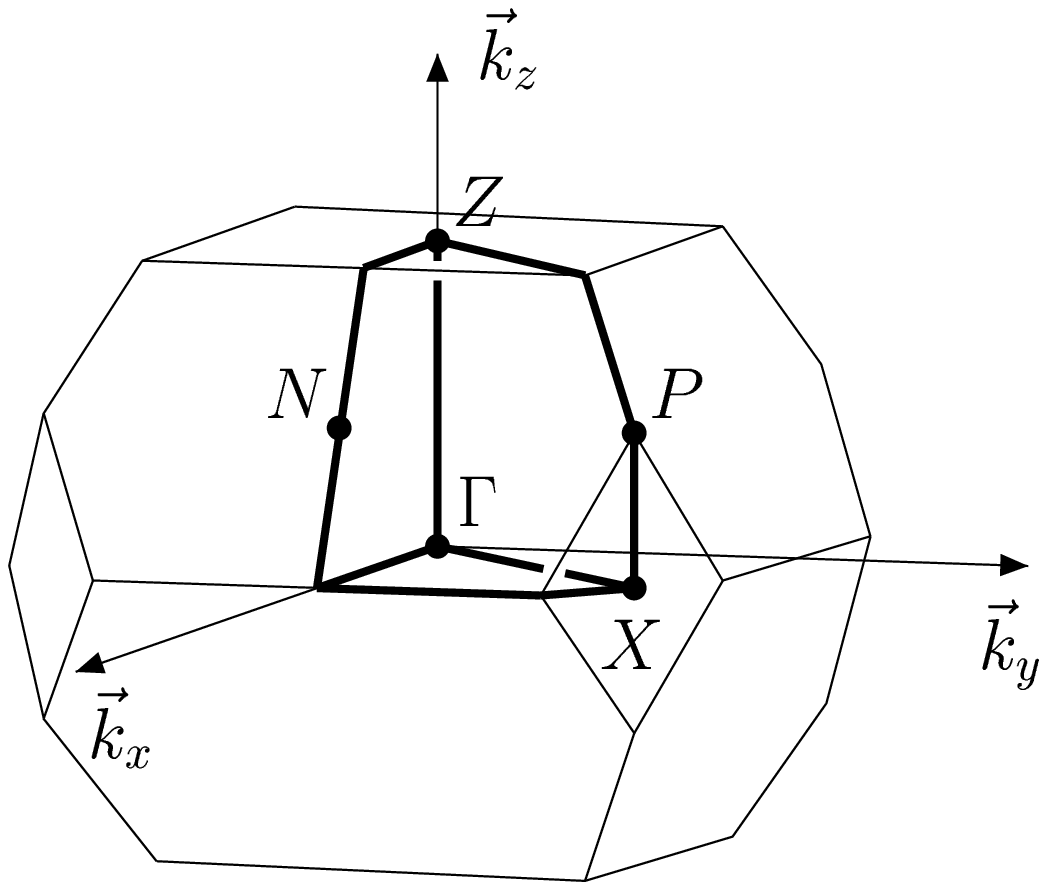}
 
   \includegraphics[width=.3\textwidth]{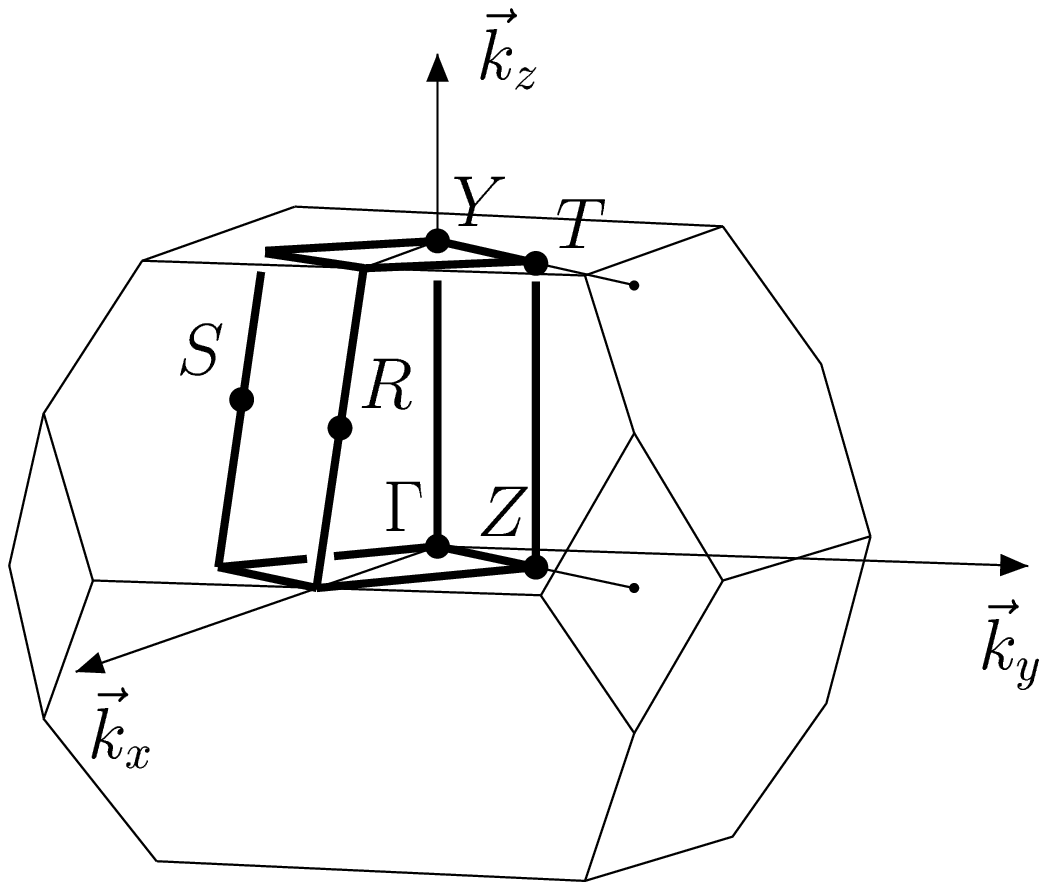}

   \includegraphics[width=.3\textwidth]{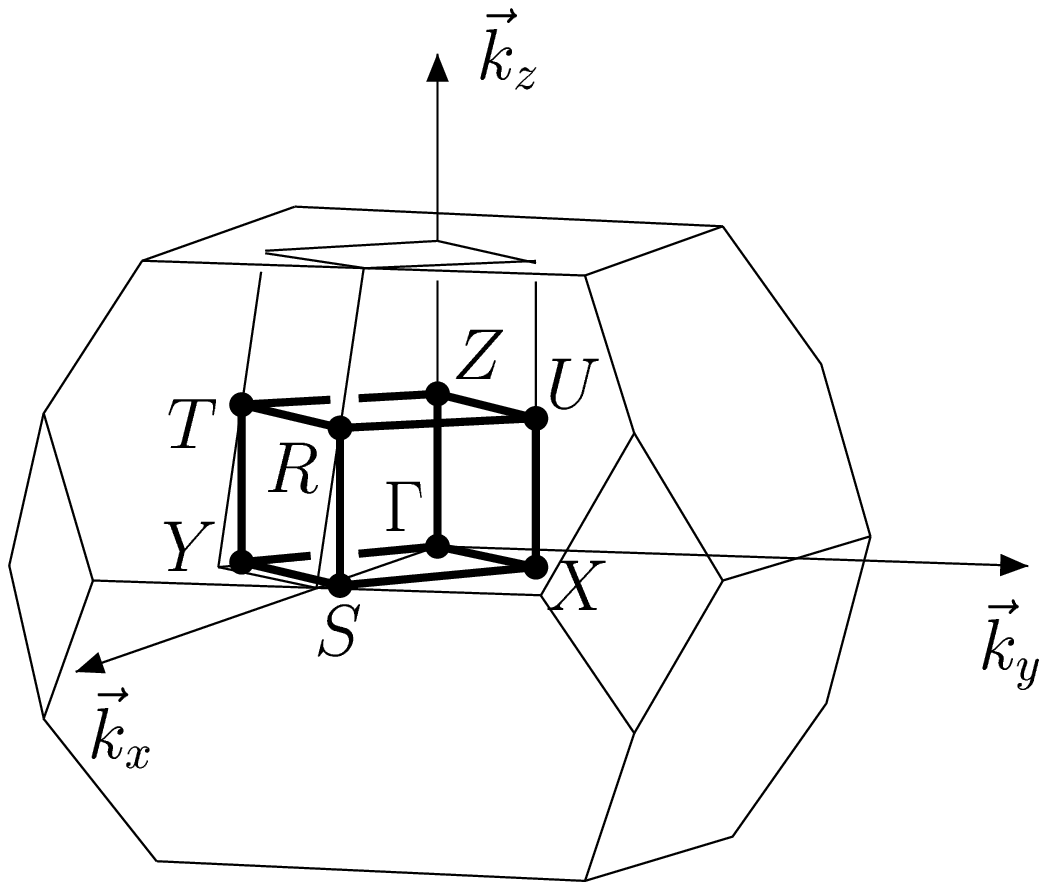}
\caption{
  Brillouin zones of La$_2$CuO$_4$.\\
  Top: Brillouin zone of the tetragonal space group $I4/mmm$ with the
  Bravais lattice $\Gamma^v_q$, as given in Fig.~3.10 (b) of
  Ref.~\protect\cite{bc}.\\ 
  Center: Basic domain of the Brillouin zone of the orthorhombic space group
  $Cmca$ with the Bravais lattice $\Gamma^b_o$ within the Brillouin zone of
  $\Gamma^v_q$.\\  
  Bottom: Basic domain of the Brillouin zone of the space group $Pccn$ of
  the antiferromagnetic phase with the Bravais lattice $\Gamma_o$ within the
  basic domain of the Brillouin zone of $\Gamma^b_o$.   
\label{fig:bz}
}
\end{figure}


In the orthorhombic phase of La$_2$CuO$_4$ the oxygen atoms are slightly
displaced from their tetragonal positions \cite{cava}, see
Fig.~\ref{fig:struktortho}.  In the Brillouin zone of this phase, the
tetragonal points $\Gamma$ and $X$ become equivalent, see Fig.~\ref{fig:bz}.
Hence, in the orthorhombic Brillouin zone we may replace the unsuited Bloch
functions at the tetragonal point $\Gamma$ by suited Bloch functions coming
from the tetragonal point $X$, if they exist. In fact, there exist
well-suited functions at the tetragonal point $X$ which are positive with
respect to the inversion and have nearly the same energy as the functions at
the tetragonal points $\Gamma$ and $Z$. These are the Bloch functions with
$X^+_2$ and $X^+_4$ symmetry, see Fig.~\ref{fig:bandstrtetra}.


\begin{figure*}[t]
\includegraphics[width=.9\textwidth,angle=0]{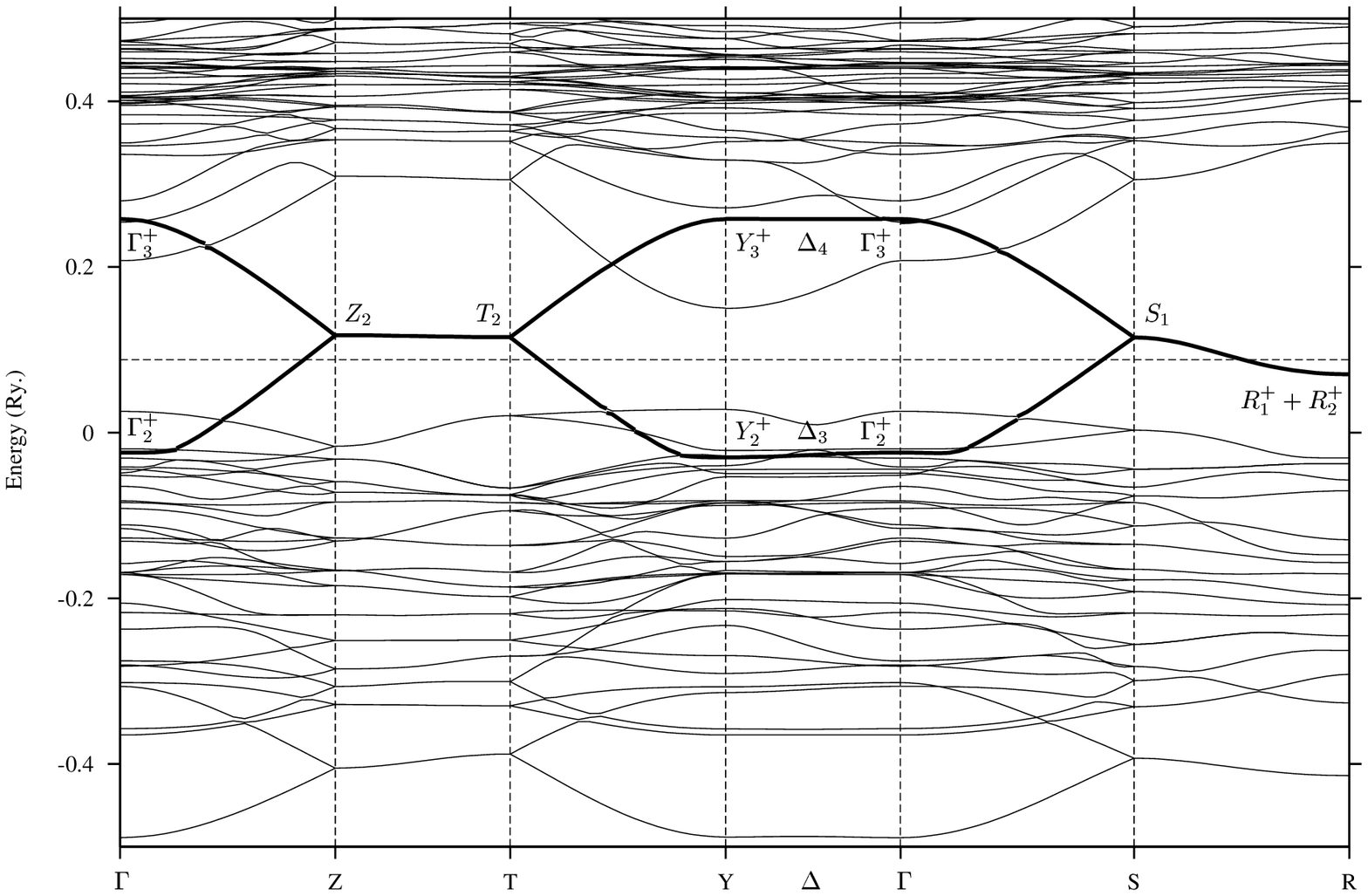}%
\caption{
  Band structure of orthorhombic La$_2$CuO$_4$ as calculated
  by Ove Jepsen \cite{jepsen}, with symmetry labels determined by the
  author. The bold line shows the ``active band'' which is proposed to be
  responsible for the physical properties of La$_2$CuO$_4$. The electrons of
  this band could occupy an energetically favorable atomiclike state by
  putting their spins antiparallel as shown in Fig. \ref{fig:struktafortho}.
  Such a spin structure with the space group $Cmca$ would not require any
  distortion of the tetragonal lattice.  However, this spin structure is not
  stable because the related magnetic group does not possess suitable
  corepresentations.  Alternatively, the electrons may occupy an atomiclike
  state either by producing the experimentally established spin structure
  with the space group $Pccn$, a ``neutral'' atomiclike state invariant
  under time-inversion 
  (and hence, without any spin structure) or a superconducting
  state, all of them being stable only in a (slightly) distorted lattice
  with the orthorhombic space group $Cmca$.
\label{fig:bandstrortho}
}
\end{figure*}


We get the ``active band'' denoted in Fig.~\ref{fig:bandstrortho} by the
bold line if we replace the tetragonal $Z^+_1$ and $\Gamma^-_2$ functions by
the tetragonal $X^+_2$ and $X^+_4$ functions. This band is proposed to be
responsible for the physical properties of La$_2$CuO$_4$.  Symmetry-adapted
Wannier functions constructed from the Bloch functions of the active band
represent localized states with orthorhombic symmetry.  However, we may
assume that these orthorhombic states do not strongly differ from states
with tetragonal symmetry, i.e., we may assume that these orthorhombic states
are states of slightly distorted tetragonal symmetry.  This is because,
first, the predominant portion of the Bloch functions of this band comes
from the Bloch functions of the initial tetragonal band \gl{edtetra}, and,
second, the new tetragonal $X$ functions have nearly the same energy as the
initial tetragonal $Z$ and $\Gamma$ functions.

The active band in the orthorhombic structure is characterized by the
representations
\begin{equation}
\Gamma^+_2 + \Gamma^+_3, Y^+_2 + Y^+_3, R^+_1 + R^+_2, T_2\text{ and }Z_2,  
\label{edafortho}
\end{equation}
where $\Gamma^+_2$ comes from the tetragonal $X^+_2$ state, $\Gamma^+_3$
comes from $X^+_3$, $Y^+_2$ comes from $X^+_4$, $Y^+_3$ comes from $X^+_3$,
too, $R^+_1 + R^+_2$ comes from $N^+_1$, and $T_2$ and $Z_2$ come from
states at the tetragonal $U$ and $\Delta$ lines, respectively, see Table
\ref{tab:faltentetraortho}.

Band \gl{edafortho} is identical to band 2 in Table \ref{tab:afbandortho}.
Hence, the Bloch functions of this band can be unitarily transformed into
optimally localizable Wannier functions symmetry-adapted to the space group
$Cmca$. The electrons of this band may lower their Coulomb energy by
occupying an atomiclike state represented by these Wannier functions.
However, the related nonadiabatic Hamiltonian $H^n$ has no longer the
symmetry of the tetragonal space group $I4/mmm$ since it commutes only with
those space group operators of the tetragonal group $I4/mmm$ which also
belong to the subgroup $Cmca$. Therefore, such an atomiclike state
represented by Wannier functions of reduced symmetry requires
a change of the symmetry of the crystal.

\subsection{Antiferromagnetic band with the orthorhombic space group
  $\bm{Cmca}$} 
\label{sec:afortho}

\begin{figure}[!]
\includegraphics[width=.3\textwidth,angle=0]{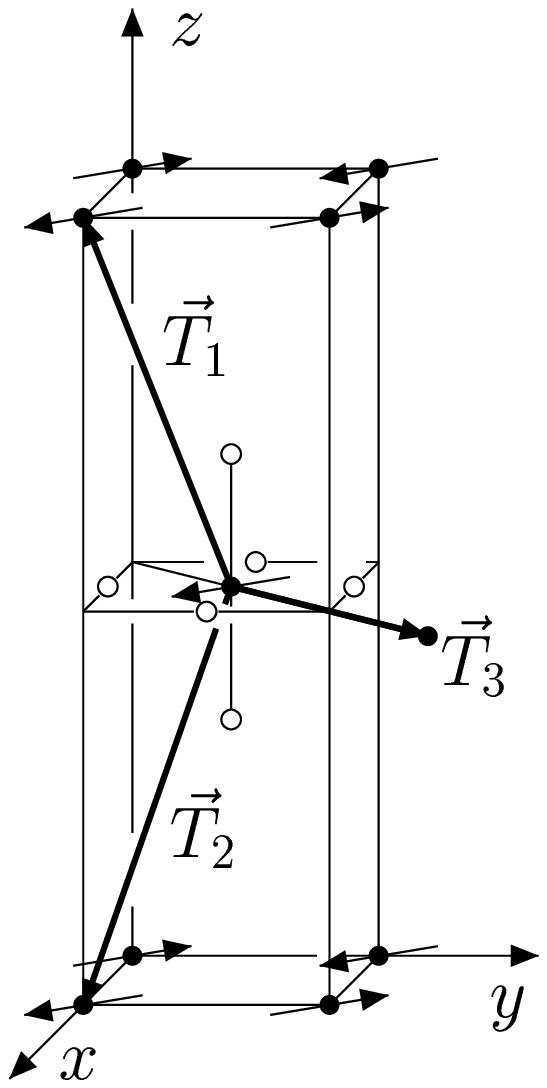}%
\caption{
  Hypothetical spin structure of La$_2$CuO$_4$ with the space group $Cmca$.
  However, this spin structure is not stable because
  the related magnetic group does not possess suitable corepresentations. 
\label{fig:struktafortho}
}
\end{figure}

The space group $Cmca$ is the space group of the antiferromagnetic structure
depicted in Fig.~\ref{fig:struktafortho}. Thus, the electrons might occupy
an atomiclike state by putting their spins antiparallel as shown in
Fig.~\ref{fig:struktafortho}. This appears to be possible, because the
Wannier functions of band 2 in Table \ref{tab:afbandortho} can be
constructed in such a way that they are symmetry-adapted to the whole
magnetic group
\begin{equation}
M = Cmca + \{K|\textstyle\frac{1}{2}\frac{1}{2}\frac{1}{2}\}Cmca,
\label{equ:mortho}
\end{equation}
where $K$ denotes the operator of time-inversion, see Appendix
\ref{sec:baender}. In this case, no spatial distortion of the tetragonal
lattice would be required, since the spin structure by itself together with
the accompanying magnetostriction represents a change of the tetragonal
symmetry.


However, within the NHM stable spin structures with the space group $Cmca$
do not exist because $M$ does not possess suitable corepresentations, see
Appendix~\ref{sec:condition}.

\subsection{Spatially distorted crystal with the orthorhombic space group
  $\bm{Cmca}$}
\label{sec:spatial}

Hence, the electrons can occupy the energetically favorable atomiclike state
only in a {\em spatially} distorted crystal. Therefore, the system produces
the experimentally established distortion with the space group $Cmca$. This
is possible because of the favorable position of the oxygen atoms: suitable
displacements of these atoms also produce a crystal with the
antiferromagnetic space group $Cmca$. Within the distorted system, there
exist three different atomiclike states.

\subsubsection{Neutral band} 
\label{sec:neutralortho}
The Wannier functions of band 2 in Table \ref{tab:afbandortho} may also be
constructed so that they are symmetry-adapted to the grey magnetic group
\begin{equation}
M = Cmca + K\!\cdot\! Cmca.
\label{equ:gmortho}
\end{equation}
Thus, in this spatially distorted crystal the conduction electrons can
occupy a ``neutral'' atomiclike state represented by these spin-independent
Wannier functions being invariant under time-inversion. I believe that this
neutral state is responsible for the phase of La$_{2-x}$Sr$_x$CuO$_4$ which
is neither antiferromagnetic nor superconducting.

\subsubsection{Antiferromagnetic band with the space group $Pccn$ of
  the experimentally established antiferromagnetic structure}
\label{sec:af}


\begin{figure}[!]
\includegraphics[width=.3\textwidth,angle=0]{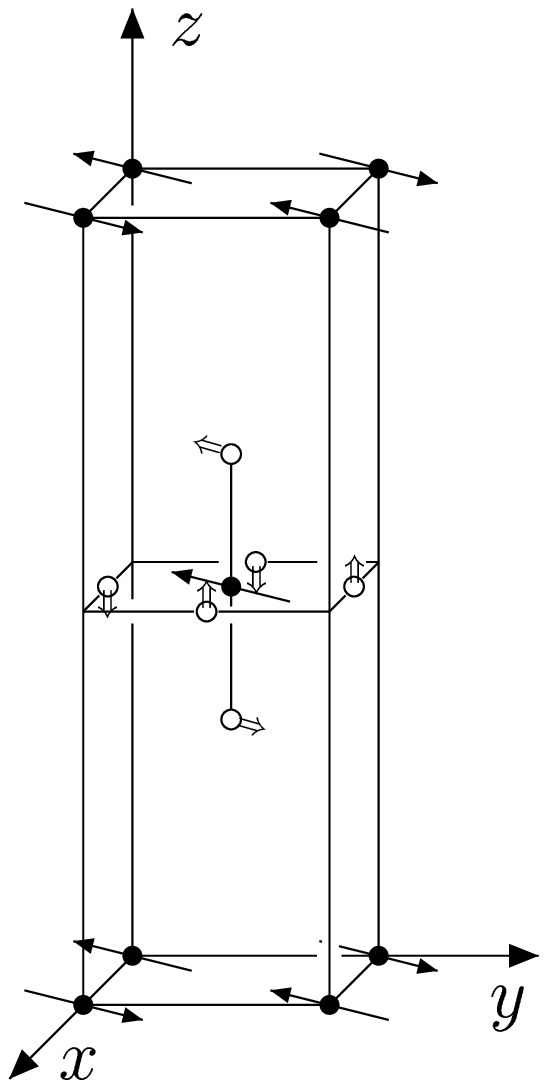}%
\caption{
  Experimentally established \cite{keimer} spin structure of La$_2$CuO$_4$
  with the space group $Pccn$. Open circles are O, solid are Cu. La is not
  shown. The arrows $\Uparrow{}$ indicate the distortions of the O atoms.
\label{fig:struktaf}
}
\end{figure}


The experimentally established \cite{keimer} antiferromagnetic structure of
La$_2$CuO$_4$ with the space group $Pccn$ is depicted in
Fig.~\ref{fig:struktaf}. Folding the active band \gl{edafortho} into the
Brillouin zone of $Pccn$ (given in Fig.~\ref{fig:bz}), we get a band
consisting of four branches which is characterized by the representations
\begin{eqnarray}
\nonumber
&\Gamma^+_1 + \Gamma^+_2 + \Gamma^+_3 + \Gamma^+_4,  
R^+_1 + R^+_2 + R^+_3 + R^+_4,\\
\nonumber 
&Y_1 + Y_2, X_1 + X_2, Z_1 + Z_2,\\
&U_1 + U_2, T_1 + T_2,\text{ and } S_1 + S_2,
\label{edaf}
\end{eqnarray}
see Table \ref{tab:faltenorthoaf}. This band is identical to band 1 in Table
\ref{tab:afband}. Hence, the Bloch functions of this band can be unitarily
transformed into optimally localizable Wannier functions symmetry-adapted to
the space group $Pccn$. Thus, within the NHM the electrons may lower their
Coulomb energy by producing the antiferromagnetic structure shown in
Fig.~\ref{fig:struktaf}. However, this spin structure is connected with the
experimentally established spatial distortion of the tetragonal lattice
because the spin structure {\em alone} does not produce a crystal with the
space group $Pccn$.


\begin{figure}[!]
\includegraphics[width=.3\textwidth,angle=0]{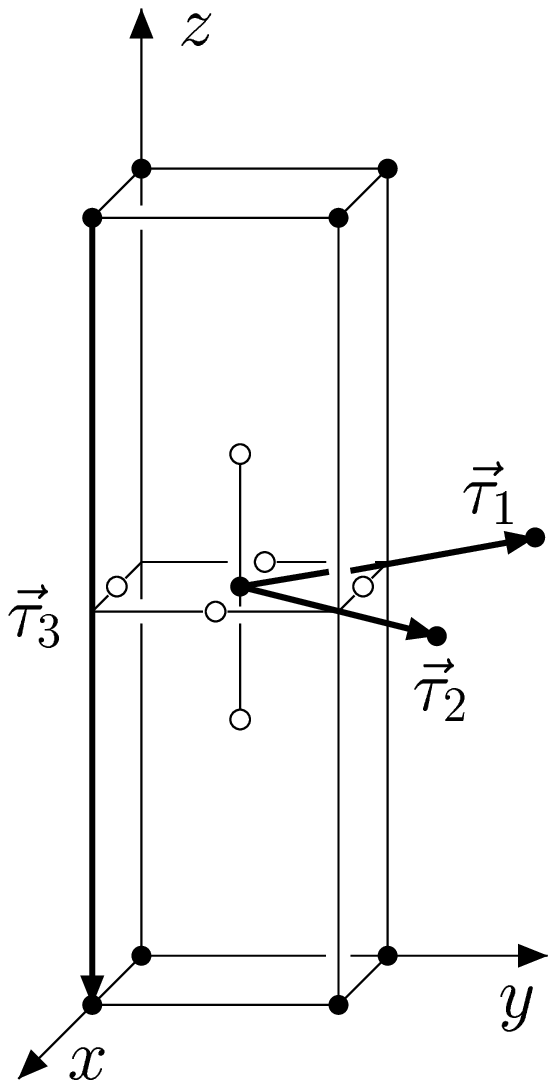}%
\caption{
  Basic translations of the Bravais lattice $\Gamma_o$ of the experimentally
  established spin structure of La$_2$CuO$_4$ shown in
  Fig. \ref{fig:struktaf}.
\label{fig:struktaftransl}
}
\end{figure}


Fig.~\ref{fig:struktaf} suggests that the related magnetic group may either
be written as 
\begin{equation}
M = Pccn + \{K|\textstyle\frac{1}{2}\frac{1}{2}0\}Pccn
\label{Mrichtig}
\end{equation}
or
\begin{equation}
M = Pccn + \{K|\textstyle\frac{1}{2}0\frac{1}{2}\}Pccn,
\label{Mfalsch}
\end{equation}
where still $K$ denotes the operator of time-inversion.

The first magnetic group \gl{Mrichtig} possesses suitable corepresentations
and, hence, spin structures with this group can be stable within the NHM,
see Appendix~\ref{sec:condition}.

\subsubsection{Superconducting band}
\label{sec:superconductivity}
Taking into account the electron spin, the active band \gl{edafortho} is
characterized by the double-valued representations
\begin{equation}
2\Gamma^+_5, 2Y^+_5, 2R^+_3 + 2R^+_4, T_3 + T_4\text{ and }Z_3 + Z_4,  
\label{zdortho}
\end{equation}
see Table \ref{tab:edzdortho}.  This band is identical to band 1 in Table
\ref{tab:slbandortho}. Hence, the Bloch functions of this band can be
unitarily transformed into optimally localizable spin-dependent Wannier
functions symmetry-adapted to the space group $Cmca$. The electrons may
lower their Coulomb energy by occupying an atomiclike state represented by
these spin-dependent Wannier functions. In this case, the conservation of
crystal spin requires a strong coupling between the electron spins and the
crystal spins of suitable boson excitations. Below a transition temperature,
this coupling leads to a superconducting state. However, this
superconducting state exists only in the space group $Cmca$ and hence, is
connected with the experimentally established spatial distortion of the
tetragonal lattice.

\section{Summary and Discussion}
\subsection{Pure La$_2$CuO$_4$}
The symmetry of the optimally localizable Wannier functions in the
conduction band of La$_2$CuO$_4$ suggests that some aspects of the physical
properties of this compound can be understood within the NHM in terms of
magnetic, neutral, and superconducting bands.

The conduction band of tetragonal La$_2$CuO$_4$ is clearly an
``antiferromagnetic'' band tending to produce the antiferromagnetic
structure with the space group $Cmca$ illustrated in
Fig.~\ref{fig:struktafortho}. The conduction electrons would be able to
occupy an atomiclike state by producing this spin structure without any
additional spatial distortion of the tetragonal crystal. However, the
nonadiabatic Hamiltonian $H^n$ does not possess antiferromagnetic {\em
  eigenstates} with the space group $Cmca$ because the related magnetic
group has no suitable corepresentations.

Thus, the system is forced to produce a {\em spatial} distortion of the
crystal so the conduction electrons can occupy the energetically favorable
atomiclike state.  We recognize the intention of the nature to stabilize the
atomiclike state with the least possible distortion: the positions of the
oxygen atoms are shifted in a subtle manner in order that the basic
translation $\vec t_3$ (see Fig.~\ref{fig:strukttetra}) in the tetragonal
lattice becomes doubled while the other two basic translations $\vec t_1$
and $\vec t_2$ still connect two lattice points. This slightly distorted
crystal has the experimentally established orthorhombic space group $Cmca$,
that is, the same space group as the spin structure in
Fig.~\ref{fig:struktafortho}. Within the distorted system, the conduction
electrons now have three possibilities to condense into an atomiclike state.

First, in the orthorhombic crystal the Bloch functions of the conduction
band can be unitarily transformed into optimally localizable Wannier
functions adapted to the symmetry of the experimentally established spin
structure (depicted in Fig.~\ref{fig:struktaf}) with the space group $Pccn$.
In this case, the related magnetic group [given in Eq.~\gl{Mrichtig}]
possesses suitable corepresentations in order that $H^n$ has eigenstates
with this spin structure. Therefore, the electrons may occupy the
energetically favorable atomiclike state by producing this antiferromagnetic
structure. However, it is the spatial orthorhombic distortion of the crystal
which stabilizes the antiferromagnetic structure.

Secondly, the orthorhombic distortion of the crystal need not be accompanied
by a spin structure. In the distorted system there exist optimally
localizable (usual) Wannier functions which are adapted to the orthorhombic
symmetry. I believe that the ``neutral'' atomiclike state represented by
these Wannier functions is responsible for the ``spin-glass'' phase
\cite{keimer} of doped La$_2$CuO$_4$ because this phase neither is
antiferromagnetic nor is superconducting.  However, the question why such a
neutral state has features of spin-glass systems \cite{sternlieb}, requires
further investigation.  Also in this case, it is the experimentally
established orthorhombic distortion which stabilizes the neutral phase.

Thirdly, in the orthorhombic system the conduction electrons may occupy an
atomiclike state represented by {\em spin-dependent} Wannier functions.  In
such an atomiclike state, the conservation of crystal spin requires a strong
coupling between the electron spins and the crystal spins of stable
crystal-spin-1 boson excitations. Below a transition temperature $T_c$,
this spin-boson interaction leads to the formation of Cooper pairs.  Also
this spin-boson phase is not stable in the tetragonal system but is
stabilized by the orthorhombic distortion.

Most likely, in any material the stable crystal-spin-1 bosons of lowest
excitation energy are coupled phonon-plasmon modes. In the isotropic
lattices of the transition elements they have dominant phonon character.
However, phonon-like excitations are not able to transport crystal-spin
angular-momenta within the two-dimensional copper-oxygen layers of
La$_2$CuO$_4$ \cite{ehtc}. Thus, in the spin-boson phase the electron spins
interact with boson excitations of dominant plasmon character.  The
relatively high energy of these plasmon-like crystal-spin-1 bosons is
proposed to be responsible for the high superconducting transition
temperature observed in the doped material.

\subsection{Doped La$_{2-x}$Sr$_x$CuO$_4$}
In doped La$_{2-x}$Sr$_x$CuO$_4$ the symmetry of the crystal is disturbed
and all the statements in the preceding section are, strictly speaking, no
longer valid. However, we may assume that for small values of $x$ the given
description of the atomiclike states holds also in the doped system because
the symmetry of the physically relevant copper-oxygen layers is not
seriously affected by the doping.

In doped La$_{2-x}$Sr$_x$CuO$_4$, all three possible atomiclike states
described in the preceding section are evidently realized: for $0 \leq x <
0.015$ the antiferromagnetic state has the lowest energy, then a neutral
phase is stable and above $x \approx 0.05$ the material becomes
superconducting \cite{keimer}. I propose a simple model which might help to
understand this $x$ dependence of the physical properties of
La$_{2-x}$Sr$_x$CuO$_4$:

At the transition from the adiabatic to the nonadiabatic system (i.e., at
the transition from a purely bandlike to the atomiclike state), the total
energy of the electron system decreases by the ``condensation energy''
\begin{equation}
\Delta E_i(x) = \Delta E^e_i - \Delta E^d_i(x),
\label{delta}
\end{equation}
where $i = 1$, 2, and 3 labels the antiferromagnetic, the neutral, and the
spin-boson phase, respectively.  (Below the transition temperature, the
latter phase is superconducting.)

$\Delta E^e_i$ stands for the electronic part of the condensation energy. It
is defined in Eqs.~(2.20) and (5.3) of Ref.~\cite{enhm} for the
superconducting and magnetic state, respectively, and may be assumed to be
independent of the Sr concentration $x$ since the physically relevant
copper-oxygen planes are not seriously affected by the doping.

Atomiclike states in La$_{2}$CuO$_4$ require an orthorhombic distortion of
the tetragonal crystal. Hence, the total condensation energy $\Delta E_i(x)$
decreases by the energy $\Delta E^d_i(x)$ required to distort the tetragonal
crystal.  The $x$-dependence of this ``distortion energy'' might be
understood on the basis of the following assumptions.

With increasing $x$ the {\em translation symmetry} of the crystal is
increasingly disturbed and the points $\Gamma$ and $X$ in the tetragonal
Brillouin zone become increasingly equivalent. As a consequence, the
orthorhombic distortion that is required for that $\Gamma$ and $X$ are
(completely) equivalent, becomes smaller with increasing $x$ (and disappears
completely for $x \approx 0.2$). Thus, for fixed $i$, the required
distortion energy $\Delta E^d_i(x)$ decreases with increasing $x$:
\begin{equation}
\Delta E^d_i(x_1) > \Delta E^d_i(x_2)\quad \text{ if }\quad x_1 < x_2. 
\label{deltafallend}
\end{equation}  

\begin{figure}[!]
\includegraphics[width=.5\textwidth,angle=0]{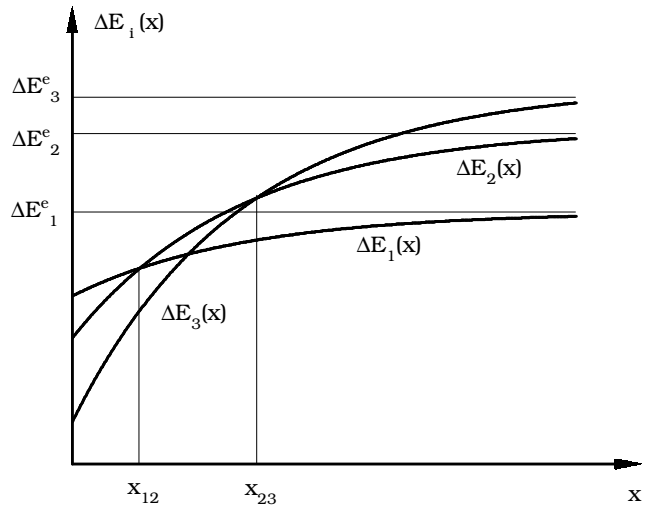}%
\caption{
  Qualitative plot of the condensation energy $\Delta E_i(x)$ [given in
  Eq.~\protect\gl{delta}] as a function of the Sr concentration $x$ within
  the simple model defined by the three assumptions
  \protect\gl{deltafallend}, \protect\gl{deltasteigend}, and
  \protect\gl{deltaxsteigend}. $i = 1$, 2, and 3 labels the
  antiferromagnetic, the neutral, and the spin-boson phase,
  respectively. The condensation 
  energy is maximum for any concentration, if the phase transitions from the
  antiferromagnetic to the neutral and from the neutral to the
  spin-boson phase occur at the Sr concentrations $x_{12}$ and
  $x_{23}$, respectively. Below a transition temperature, the spin-boson
  phase is superconducting.
\label{fig:deltae}
}
\end{figure}

On the basis of this assumption, the experimental results may be interpreted
by assuming that, first, the purely electronic condensation energy is
maximum for the spin-boson and minimum for the antiferromagnetic state,
\begin{equation}
\Delta E^e_1 < \Delta E^e_2 < \Delta E^e_3, 
\label{deltasteigend}
\end{equation}  
and that, second, for fixed $x$, the required distortion energy is also
maximum for the spin-boson and minimum for the antiferromagnetic state,
\begin{equation}
\Delta E^d_1(x) < \Delta E^d_2(x) < \Delta E^d_3(x). 
\label{deltaxsteigend}
\end{equation}  
Under these assumptions, the system undergoes the phase transitions in order
that the condensation energy is maximum at any Sr concentration $x$, see
Fig.~\ref{fig:deltae}.


The first assumption \gl{deltasteigend} is in accordance with my
observations on the elemental metals, see Sec.~V of Ref.~\cite{ef}. The
second assumption is corroborated by the fact that a spin structure alone
(together with the related magnetostriction) represents a distortion of the
lattice without any spatial distortion. Hence, it is conceivable that the
required {\em spatial} distortion of the tetragonal crystal is smallest in
the antiferromagnetic state.

\appendix
\section{stable magnetic states}
\label{sec:condition}
Consider a magnetic state $|m\rangle$ and the related magnetic
group 
\begin{equation}
M = H + AH,
\label{mgroup}
\end{equation}
where $H$ denotes the space group and $A$ stands for any anti-unitary
element of $M$. In the case of the spin structure depicted in Fig.\
\ref{fig:struktafortho}, we have $H = Cmca$ and may choose $A =
\{K|\frac{1}{2}\frac{1}{2}\frac{1}{2}\}$ with $K$ denoting the operator of
time-inversion.  

Within the NHM, both the magnetic state $|m\rangle$ and the time-inverted
state
\[
\overline{|m}\rangle = K|m\rangle   
\]
are {\em eigenstates} of the nonadiabatic Hamiltonian $H^n$. Therefore,
standard group-theoretical methods can be applied to magnetic states, too.
A {\em stable} magnetic state $|m\rangle$ complies with two conditions:
\begin{itemize}
\item 
$|m\rangle$ is basis function of a one-dimensional corepresentation
of $M$;
\item $|m\rangle$ and the time-inverted state $\overline{|m}\rangle$
are basis functions of a two-dimensional {\em irreducible} corepresentation
of the gray magnetic group
\begin{equation}
\overline{M} = M + KM,
\label{graymgroup}
\end{equation}
\end{itemize}
see Sec.~III.C of Ref.\ \cite{ea}.

Consequently, a stable magnetic state can exist in the space group $H$ if
$H$ has at least one one-dimensional single-valued representation $R$
following
\begin{itemize}
\item 
case (a) with respect to the magnetic group $H + AH$ and
\item 
case (c) with respect to the magnetic group \mbox{$H + KH$.}
\end{itemize}
The cases (a) and (c) are defined in Eqs.\ (7.3.45) and (7.3.47),
respectively, in the textbook of Bradley and Cracknell \cite{bc}. The
irreducible corepresentation derived from $R$ stays one-dimensional in case
(a) and becomes two-dimensional in case (c).

\subsection{Space group $\bm{Cmca}$}
Table \ref{tab:edortho} lists all the irreducible representations of the
space group $H = Cmca$ of orthorhombic La$_2$CuO$_4$. The second and third
column of this Table specify whether the corepresentation of $H+KH$ and 
$H+\{K|\frac{1}{2}\frac{1}{2}\frac{1}{2}\}H$, respectively, derived from the
given representation $R$ follows case (a) or case (c).

The only one-dimensional representations which apparently comply with the
above condition for a stable antiferromagnetic state are the small
representations at point $R$. However, the little group at $R$ is not the
complete space group. Therefore, a stable antiferromagnetic state with the
space group $Cmca$ does not exist.

\subsection{Space group $\bm{Pccn}$}

Table \ref{tab:edaf} lists all the irreducible representations of the space
group $H = Pccn$. The second, third, and fourth column of this table specify
whether the corepresentation of $H + KH, H +
\{K|\frac{1}{2}\frac{1}{2}0\}H,\mbox{ and } H +
\{K|\frac{1}{2}0\frac{1}{2}\}H$ [cf.~Eqs.~\gl{Mrichtig} and \gl{Mfalsch}],
respectively, derived from the given representation $R$ follows case (a) or
case (c).

For the group $H + \{K|\frac{1}{2}0\frac{1}{2}\}H$ there do not exist
suitable representations. However, for the group $H +
\{K|\frac{1}{2}\frac{1}{2}0\}H$ all the small representations at point $R$
comply with the above condition for a stable antiferromagnetic state. Since
now the little group at $R$ includes all the space-group elements, a stable
antiferromagnetic state with the space group $Pccn$ is possible.

\section{Optimally localizable and symmetry-adapted Wannier functions} 
\label{sec:baender}
\subsection{Usual (spin-independent) Wannier functions}
Consider a subgroup $H$ of the space group $I4/mmm$ of La$_2$CuO$_4$ and
assume the energy bands (or a part of the energy bands) of La$_2$CuO$_4$ to
be folded into the Brillouin zone $B_H$ of $H$. Let be $\mu$ the number of
Cu atoms in the unit cell of $H$ and consider $\mu$ single energy bands in
$B_H$. Assume the Bloch functions $\varphi_{\vec kq}(\vec r)$ (with the wave
vector $\vec k$) of this ``energy band with $\mu$ branches'' to be unitarily
transformed into Wannier functions
\begin{equation}
  \label{eq:wf}
w_i(\vec r - \vec T - \vec\rho_i) = \frac{1}{\sqrt{\nu}}\sum^{BZ}_{\vec
  k}\sum_{q = 1}^{\mu}e^{-i\vec k (\vec T + \vec\rho_i)} g_{iq}(\vec k
  )\varphi_{\vec kq}(\vec r)   
\end{equation}
centered at the Cu atoms. The first sum in this equation runs over the $\nu$
vectors $\vec k$ of the first Brillouin zone (BZ), the second sum runs over
the $\mu$ branches, $\vec T$ and $\vec\rho_i$ denote the vectors of the
Bravais lattice and the positions of the Cu atoms within the unit cell,
respectively, and the coefficients $g_{iq}(\vec k)$ are the elements of a
unitary matrix ${\bf g}(\vec k)$,
\begin{equation}
  \label{eq:gk}
{\bf g}^{-1}(\vec k) = {\bf g}^{\dagger}(\vec k).  
\end{equation}

The Wannier functions are called ``optimally localizable'' if the Bloch-like
functions
\begin{equation}
  \label{eq:26}
\widetilde\varphi_{\vec ki}(\vec r) = \sum_{q = 1}^{\mu}g_{iq}(\vec k
  )\varphi_{\vec kq}(\vec r)     
\end{equation}
are (for fixed $\vec r$) continuous functions of $\vec k$ everywhere in the
reciprocal space. Fortunately, the question whether or not the functions
$\widetilde\varphi_{\vec ki}(\vec r)$ vary continuously through the planes
of symmetry within or on the surface of the Brillouin zone, can be treated
by group theory \cite{ew1,ew2}. The main result is the relatively simple
criterion given in Eq.~\gl{eq:31}.

The Wannier functions are called ``symmetry-adapted to $H$'' if
they satisfy the equation
\begin{equation}
  \label{eq:hadapted}
P(a)w_i(\vec r - \vec T - \vec\rho_i) = \sum_{j =
  1}^{\mu}D_{ji}(\alpha )w_j(\vec r - \alpha\vec T - \vec\rho_j) 
\end{equation}
by application of the operators $P(a)$ of $H$,
\begin{equation}
  \label{eq:ah}
a = \{\alpha |\vec t\}\in H,
\end{equation}
see Appendix A of Ref.~\cite{enhm}. The matrices ${\bf D}_0(\alpha ) =
[D_{ji}(\alpha )]$ form a (reducible or irreducible) single-valued
representation $D_0$ of the point group $H_0$ of $H$. The sum on the right
hand side of Eq.~\gl{eq:hadapted} consists of one summand only because there
is one Wannier function at each atom. Hence, the matrices ${\bf D}_0(\alpha
)$ have only one non-vanishing element, say $d_{ji}(\alpha )$, in each row
and each column.  The non-vanishing diagonal elements $d_{jj}(\alpha )$ are
equal to $\chi_0(\alpha )$ and the other non-vanishing elements may also be
chosen to be equal to $\chi_0(\alpha )$, so that we have
\begin{equation}
\label{eq:dji}
d_{ji}(\alpha ) = \chi_0(\alpha )
\end{equation}
for all the non-vanishing elements of ${\bf D}_0(\alpha )$, where
$\chi_0(\alpha )$ stands for the character of the one-dimensional
representation $d_0$ of $H_0$ used in the following Eq.~\gl{eq:31}.

Suppose the symmetry of the Bloch functions in the Brillouin zone $B_H$ of
$H$ to be determined by the compatibility relations given in the Tables
\ref{tab:faltentetraortho} - \ref{tab:faltenorthoaf}. At each point of
symmetry $P_{\vec k}$ of $B_H$ the Bloch functions of the $\mu$ branches of
the band belong to a $\mu$-dimensional single-valued small representation
$D_{\vec k}$ of the little group $H_{\vec k}$ of $P_{\vec k}$.  $D_{\vec k}$
may be irreducible or is the direct sum over small irreducible
representations of $H_{\vec k}$.

The coefficients $g_{iq}(\vec k)$ can be chosen so that the Wannier
functions in Eq.~\gl{eq:wf} are optimally localizable and symmetry-adapted
to $H$ if and only if the equation
\begin{equation}
  \label{eq:31}
\chi_{\vec k}(a)= \chi_0(\alpha )e^{-i\alpha\vec k\cdot\vec t}~
\sum_{i = 1}^{\mu}n_{i}(a)e^{-i\vec \varrho_{i}\cdot (\vec k - \alpha\vec k)}
\end{equation}
is satisfied at each $P_{\vec k}$ of $B_H$ and for each element \mbox{$a =
  \{\alpha|\vec t\}$} in the little group $H_{\vec k}$.  $\chi_{\vec k}(a)$
denotes the character of the representation $D_{\vec k}$, $\chi_0(\alpha)$
stands for the character of any one-dimensional single-valued representation
$d_0$ of the point group $H_0$ of $H$, and
\begin{equation}
  \label{eq:32}
n_{i}(a) = \left\{
\begin{array}{r l}
1 & \mbox{if } \alpha\vec \varrho_{i} + \vec t = \vec \varrho_{i} + \vec T\\
0 & \mbox{else}
\end{array}
\right.
\end{equation}
with $\vec T$ being a vector of the Bravais lattice. 
This Eq.~\gl{eq:31} is
derived from Eqs.~(1.8) and (4.28) of Ref.~\cite{ew1} (see also Sec.~3
of Ref.~\cite{ew2}). It is satisfied for all $a \in H_{\vec k}$ if it
is satisfied for one element of each class of $H_{\vec k}$.

\subsection{Spin-dependent Wannier functions}
The Wannier functions given in Eq.~\gl{eq:wf} become spin-dependent if the
Bloch functions $\varphi_{\vec kq}(\vec r)$ in this equation are allowed to
have $\vec k$-dependent spin-directions, see Eqs.~(A21) and (A22) in
Ref.~\cite{enhm}.  Also energy bands with optimally localizable and
symmetry-adapted spin-dependent Wannier functions situated at the Cu atoms
follow Eq.~\gl{eq:31}. In this case, however, $D_{\vec k}$ is a
$2\mu$-dimensional double-valued representation and $\chi_0(\alpha)$ denotes
the character of any two-dimensional double-valued representation $d_0$ of
$H_0$.

\subsection{Magnetic groups}
The Wannier functions given in Eq.~\gl{eq:wf} are called symmetry-adapted
to the antiferromagnetic group
\begin{equation}
\label{eq:m}
M = H + \{K|\vec\tau\}H
\end{equation}
if the equation 
\begin{equation}
  \label{eq:madapted}
KP\big(\{E|\vec\tau\}\big)w_i(\vec r - \vec T - \vec\rho_i) = \sum_{j =
  1}^{\mu}N_{ji}w_j(\vec r - \vec T - \vec\rho_j) 
\end{equation}
is satisfied in addition to Eq.~\gl{eq:hadapted}, where $K$ still denotes
the operator of time-inversion. Since there is one Wannier function at each
atom, also the matrix ${\bf N} = [N_{ji}]$ has only one non-vanishing
element, say $n_{ji}$, in each row and each column which may be put equal to
one,
\begin{equation}
  \label{eq:3}
  n_{ji} = 1.
\end{equation}
Suppose that in Eq.~\gl{eq:madapted} we obtain the Wannier function $w_j$ by
application of the operator $KP\big(\{E|\vec\tau\}\big)$ on the function
$w_i$. Then conversely we obtain the function $w_i$ applying this
operator on $w_j$. This is because $2\tau$ is a vector of the Bravais
lattice,
\begin{equation}
  \label{eq:8}
  2\vec\tau = \vec T,
\end{equation}
and hence from $\vec\rho_i + \vec\tau = \vec\rho_j + \vec T_1$ it follows
$\vec\rho_j + \vec\tau = \vec\rho_i + \vec T_2$, where $\vec T_1$ and $\vec
T_2$ denote vectors of the Bravais lattice. Thus, the matrix ${\bf N}$ is
symmetric,
\begin{equation}
  \label{eq:7}
  n_{ji} = n_{ij}.
\end{equation}

${\bf N}$ is the matrix representing $K$ in the corepresentation of the
point group
\begin{equation}
\label{pmg0}
M_0 = H_0 + KH_0  
\end{equation}
of $M$ which is derived from the representation $D_0$ of $H_0$ given in
Eq.~\gl{eq:hadapted}. Hence, ${\bf N}$ must satisfy the equations
\begin{equation}
  \label{eq:5}
  {\bf NN}^* = \bm{1}
\end{equation}
and
\begin{equation}
  \label{eq:1}
  {\bf D}_0(\alpha ) = {\bf ND}^*_0(\alpha ){\bf N}^{-1} \text{ for } \alpha
  \in H_0, 
\end{equation}
see Eq.~(7.3.45) of Ref.~\cite{bc}. The matrices ${\bf D}_0(\alpha )$ are
defined in Eq.~\gl{eq:hadapted}. [In Ref.~\cite{bc} Eq.~(7.3.45) is derived
for irreducible representations. However, this derivation in Sec.~7.3 of
Ref.~\cite{bc} shows that Eq.~(7.3.45) holds for reducible representations,
too, if Eq.~\gl{eq:5} is satisfied.]

As a consequence of Eqs.~\gl{eq:3} and \gl{eq:7}, the first Eq.~\gl{eq:5} is
satisfied (and hence ${\bf N} = {\bf N}^* = {\bf N}^{-1}$).  The validity of
the second Eq.~\gl{eq:1} must be examined for each given magnetic group.

Assume Eq.~\gl{eq:31} to be valid in a given energy band. In Ref.~\cite{ew3}
I have shown that then the coefficients $g_{iq}(\vec k)$ [in Eq.~\gl{eq:wf}]
can be chosen so that the Wannier functions of this band are
symmetry-adapted to the whole magnetic group $M$ if the matrix ${\bf N}$
satisfies Eq.~\gl{eq:1} and the equation
\begin{equation}
  \label{eq:2}
  {\bf S}^*(-\vec K) = {\bf NS}(\vec K){\bf N}^{-1}\cdot e^{i\vec
  K\cdot\tau}  
\end{equation}
for any vector $\vec K$ of the reciprocal lattice, where ${\bf S}(\vec K)$
is a 
$\mu$-dimensional diagonal matrix with the elements
\begin{equation}
  \label{eq:4}
  s_{jj} = e^{i\vec K\cdot\vec\rho_j}.
\end{equation}
Using Eqs.~\gl{eq:3}, \gl{eq:8}, and \gl{eq:7} it can be verified that the
matrix ${\bf N}$ defined by Eq.~\gl{eq:madapted} satisfies Eq.~\gl{eq:2}.
Thus, Eq.~\gl{eq:1} is the only additional condition for the existence of
optimally localizable Wannier functions that are symmetry-adapted to the
entire magnetic group $M$. This equation is satisfied for all the
antiferromagnetic bands considered in this paper.

\section{Description of the space groups}
\subsection{Tetragonal L\lowercase{a}$_2$C\lowercase{u}O$_4$}
Fig.~\ref{fig:strukttetra} shows the basic translations of the space group
$I4/mmm$ of tetragonal La$_2$CuO$_4$ and the Brillouin zone is drawn in
Fig.~\ref{fig:bz} (top).

The point group $D_{4h}$ consists of 16 elements,
\begin{eqnarray}
\nonumber
D_{4h} &=& \{E, C^+_{4z}, C^-_{4z}, C_{2z}, C_{2x}, C_{2y}, C_{2a}, 
C_{2b}, I, \\
\nonumber
&&S^-_{4z} = IC^+_{4z}, S^+_{4z} = IC^-_{4z}, \sigma_z = IC_{2z},
\sigma_x = IC_{2x},\\
\label{tetrael}
\nonumber
&&\sigma_y = IC_{2y}, \sigma_{da} = IC_{2a}, \sigma_{db} =
IC_{2b}\},\\
\end{eqnarray}
representing the identity operation, anti-clockwise and clockwise rotation
through $\pi /2$ radians about the $z$ axis, rotation through $\pi$ radians
about the $z$, $x$ and $y$ axes, rotation through $\pi$ radians
about the $a$ and $b$ axes drawn in Fig.~\ref{fig:strukttetra}, and
inversion. The following elements are products of the inversion with the
former elements.  

The space group $I4/mmm$ is symmorphic. Hence, all the point group elements
are elements of $I4/mmm$, too.


\subsection{Orthorhombic L\lowercase{a}$_2$C\lowercase{u}O$_4$}
The space group $Cmca$ of orthorhombic La$_2$CuO$_4$ has the basic
translations shown in Fig.~\ref{fig:struktortho}. The basic domain of the
Brillouin zone of $Cmca$ is drawn in Fig.~\ref{fig:bz} (center).

Within the coordinate system given in Fig.~\ref{fig:struktortho} its point
group is
\begin{equation}
D_{2h} = \{E, C_{2z}, C_{2a}, C_{2b}, I, \sigma_{z}, \sigma_{da},
 \sigma_{db}\}. 
\label{d2h} 
\end{equation}
In the space group $Cmca$ the 4
elements      
 \begin{equation}
C_{2z}, C_{2a}, \sigma_{z}, \text{ and }\sigma_{da}
\label{orthoelnp} 
\end{equation}
are associated with the nonprimitive translation
$$
({\textstyle\frac{1}{2}\frac{1}{2}\frac{1}{2}}).
$$


\subsection{Antiferromagnetic L\lowercase{a}$_2$C\lowercase{u}O$_4$}
Fig.~\ref{fig:struktaf} shows the spin structure of La$_2$CuO$_4$ as
determined by Keimer et al. \cite{keimer}. The space group is $Pccn$ with
the Bravais lattice $\Gamma_o$. Fig.~\ref{fig:struktaftransl} gives the
basic translations as used in this paper, and the basic domain of the
Brillouin zone is drawn in Fig.~\ref{fig:bz} (bottom).

The point group of the antiferromagnetic structure is also the group
$D_{2h}$ given in Eq.\ \gl{d2h}.  In the space group $Pccn$ the 2 elements
 \begin{equation}
C_{2z}\text{ and }\sigma_{z}
\label{afelnp1} 
\end{equation}
are associated with the nonprimitive translation
$$
({\textstyle\frac{1}{2}\frac{1}{2}0}),
$$
the elements
 \begin{equation}
C_{2a}\text{ and }\sigma_{da}
\label{afelnp2} 
\end{equation}
are associated with 
$$
(0{\textstyle\frac{1}{2}\frac{1}{2}}),
$$
and the elements
\begin{equation}
C_{2b}\text{ and }\sigma_{db}
\label{afelnp3} 
\end{equation}
are associated with
$$
({\textstyle\frac{1}{2}0\frac{1}{2}}).
$$
\section{Tables}
\label{sec:tables}
\subsection{Irreducible representations}
The irreducible representations given in Tables \ref{tab:edtetra} -
\ref{tab:zdortho} (on the following pages) are determined by means of Tables
5.7 and 6.13 in the textbook of Bradley and Cracknell \cite{bc}, with the
notations of tables 5.8 and 6.14 in this book. The cases (a), (b), and (c)
of the corepresentations follow Eqs.~(7.3.45), (7.3.46), and (7.3.47) and
are determined by Eq.~(7.3.51) of Ref.~\cite{bc}.

\subsection{Compatibility tables}
Tables \ref{tab:faltentetraortho} - \ref{tab:faltenorthoaf} list
compatibility relations between the points of symmetry in the Brillouin zone
of a space group $G$ and the related points of symmetry in the Brillouin
zone of a subgroup $H$ of $G$. These tables are determined in the way
described in great detail in Ref.~\cite{eabf}. The compatibility relations
between single-valued and double-valued representations given in Tables
\ref{tab:edzdtetra} and \ref{tab:edzdortho} are determined by standard
group-theoretical methods.

\subsection{Antiferromagnetic, neutral, and superconducting bands} 
Tables \ref{tab:wftetra}, \ref{tab:afbandortho}, \ref{tab:afbandcr}, and
\ref{tab:afband} list all the single-valued representations $D_{\vec k}$
with characters $\chi_{\vec k}(a)$ satisfying Eq.~\gl{eq:31} in the space
groups $H = I4/mmm, Cmca, P4/mnc, \text{ and } Pccn$, respectively.
Different bands belong to the character $\chi_0(\alpha)$ of different
one-dimensional representations $d_0$ of $H_0$.  Tables
\ref{tab:slbandtetra} and \ref{tab:slbandortho} list all the double-valued
representations satisfying this equation with $H = I4/mmm \text{ and }
Cmca$, respectively. In this case, however, $D_{\vec k}$ is a
$2\mu$-dimensional double-valued representation and $\chi_0(\alpha)$ denotes
the character of any two-dimensional double-valued representation $d_0$ of
$H_0$.

\acknowledgements{%
  I am very indebted to Ove Jepsen for providing me with all the data I
  needed to determine the symmetry of the Bloch functions in the band
  structure of La$_2$CuO$_4$. I thank Ernst Helmut Brandt for critical
  comments on the manuscript and continuing encouragement.}

\begin{table*}[p]
\caption{
  Characters of the small (allowed) single-valued irreducible
  representations of the space 
  group $I4/mmm = D^{17}_{4h}$ (139) of tetragonal La$_2$CuO$_4$ in the
  notation of Table 5.8 in the textbook of Bradley and Cracknell
  \protect\cite{bc}. 
\label{tab:edtetra}
}
\begin{tabular}[t]{cccccccccccc}
\multicolumn{12}{c}{$\Gamma (000)$ and $Z (\frac{1}{2}\frac{1}{2}\overline{\frac{1}{2}})$}\\
& & $$ & $$ & $C^-_{4z}$ & $C_{2y}$ & $C_{2b}$ & $$ & $$ & $S^+_{4z}$ & $\sigma_y$ & $\sigma_{db}$\\
& & $E$ & $C_{2z}$ & $C^+_{4z}$ & $C_{2x}$ & $C_{2a}$ & $I$ & $\sigma_z$ & $S^-_{4z}$ & $\sigma_x$ & $\sigma_{da}$\\
\hline
$\Gamma^+_1$ & $Z^+_1$ & 1 & 1 & 1 & 1 & 1 & 1 & 1 & 1 & 1 & 1\\
$\Gamma^+_2$ & $Z^+_2$ & 1 & 1 & 1 & -1 & -1 & 1 & 1 & 1 & -1 & -1\\
$\Gamma^+_3$ & $Z^+_3$ & 1 & 1 & -1 & 1 & -1 & 1 & 1 & -1 & 1 & -1\\
$\Gamma^+_4$ & $Z^+_4$ & 1 & 1 & -1 & -1 & 1 & 1 & 1 & -1 & -1 & 1\\
$\Gamma^+_5$ & $Z^+_5$ & 2 & -2 & 0 & 0 & 0 & 2 & -2 & 0 & 0 & 0\\
$\Gamma^-_1$ & $Z^-_1$ & 1 & 1 & 1 & 1 & 1 & -1 & -1 & -1 & -1 & -1\\
$\Gamma^-_2$ & $Z^-_2$ & 1 & 1 & 1 & -1 & -1 & -1 & -1 & -1 & 1 & 1\\
$\Gamma^-_3$ & $Z^-_3$ & 1 & 1 & -1 & 1 & -1 & -1 & -1 & 1 & -1 & 1\\
$\Gamma^-_4$ & $Z^-_4$ & 1 & 1 & -1 & -1 & 1 & -1 & -1 & 1 & 1 & -1\\
$\Gamma^-_5$ & $Z^-_5$ & 2 & -2 & 0 & 0 & 0 & -2 & 2 & 0 & 0 & 0\\
\hline\\
\end{tabular}\\
\begin{tabular}[t]{ccccc}
\multicolumn{5}{c}{$N (0\frac{1}{2}0)$}\\
 & $E$ & $C_{2y}$ & $I$ & $\sigma_y$\\
\hline
$N^+_1$ & 1 & 1 & 1 & 1\\
$N^-_1$ & 1 & 1 & -1 & -1\\
$N^+_2$ & 1 & -1 & 1 & -1\\
$N^-_2$ & 1 & -1 & -1 & 1\\
\hline\\
\end{tabular}\\
\begin{tabular}[t]{ccccccccc}
\multicolumn{9}{c}{$X (00\frac{1}{2})$}\\
 & $E$ & $C_{2z}$ & $C_{2a}$ & $C_{2b}$ & $I$ & $\sigma_z$ & $\sigma_{da}$ & $\sigma_{db}$\\
\hline
$X^+_1$ & 1 & 1 & 1 & 1 & 1 & 1 & 1 & 1\\
$X^+_2$ & 1 & -1 & 1 & -1 & 1 & -1 & 1 & -1\\
$X^+_3$ & 1 & 1 & -1 & -1 & 1 & 1 & -1 & -1\\
$X^+_4$ & 1 & -1 & -1 & 1 & 1 & -1 & -1 & 1\\
$X^-_1$ & 1 & 1 & 1 & 1 & -1 & -1 & -1 & -1\\
$X^-_2$ & 1 & -1 & 1 & -1 & -1 & 1 & -1 & 1\\
$X^-_3$ & 1 & 1 & -1 & -1 & -1 & -1 & 1 & 1\\
$X^-_4$ & 1 & -1 & -1 & 1 & -1 & 1 & 1 & -1\\
\hline\\
\end{tabular}\\
\begin{tabular}[t]{cccccc}
\multicolumn{6}{c}{$P (\frac{1}{4}\frac{1}{4}\frac{1}{4})$}\\
 & $$ & $$ & $S^-_{4z}$ & $C_{2y}$ & $\sigma_{da}$\\
 & $E$ & $C_{2z}$ & $S^+_{4z}$ & $C_{2x}$ & $\sigma_{db}$\\
\hline
$P_1$ & 1 & 1 & 1 & 1 & 1\\
$P_2$ & 1 & 1 & 1 & -1 & -1\\
$P_3$ & 1 & 1 & -1 & 1 & -1\\
$P_4$ & 1 & 1 & -1 & -1 & 1\\
$P_5$ & 2 & -2 & 0 & 0 & 0\\
\hline\\
\end{tabular}
\end{table*}
\begin{table*}[p]
\caption{
  Characters of the small (allowed) single-valued irreducible
  representations of the space 
  group $H = Cmca = D^{18}_{2h}$ (64) of the orthorhombic phase of
  La$_2$CuO$_4$ in the notation of Table 5.8 in the textbook of Bradley and
  Cracknell \protect\cite{bc}. $K$ stands for the operator of 
  time-inversion. The 
  first two elements in the first row define two magnetic groups $M_1 = H +
  KH \quad\mbox{ and} \quad M_2 = H +\{K|\frac{1}{2}\frac{1}{2}\frac{1}{2}\}
  H$. The second and third column of the table specify whether the
  corepresentations of $M_1$ and $M_2$, respectively, derived from the given
  representations $R_i$ of $H$ follow case $(a)$, case $(b)$, or case $(c)$
  when they are given by Eqs.  (7.3.45-47) of Ref.~\protect\cite{bc}. In
  case (a), there 
  is no change in the degeneracy of $R_i$, and in case (c) the degeneracy
  becomes doubled. 
\label{tab:edortho}
}
\begin{tabular}[t]{cccccccccccc}
\multicolumn{12}{c}{$\Gamma(000)$ and $Y(\overline{\frac{1}{2}}\frac{1}{2}0)$}\\
& & $K$ & $\{K|\frac{1}{2}\frac{1}{2}\frac{1}{2}\}$ & $\{E|000\}$ & $\{C_{2a}|\frac{1}{2}\frac{1}{2}\frac{1}{2}\}$ & $\{C_{2z}|\frac{1}{2}\frac{1}{2}\frac{1}{2}\}$ & $\{C_{2b}|000\}$ & $\{I|000\}$ & $\{\sigma_{da}|\frac{1}{2}\frac{1}{2}\frac{1}{2}\}$ & $\{\sigma_z|\frac{1}{2}\frac{1}{2}\frac{1}{2}\}$ & $\{\sigma_{db}|000\}$\\
\hline
$\Gamma^+_1$ & $Y^+_1$ & (a) & (a) & 1 & 1 & 1 & 1 & 1 & 1 & 1 & 1\\
$\Gamma^+_2$ & $Y^+_2$ & (a) & (a) & 1 & -1 & 1 & -1 & 1 & -1 & 1 & -1\\
$\Gamma^+_3$ & $Y^+_3$ & (a) & (a) & 1 & 1 & -1 & -1 & 1 & 1 & -1 & -1\\
$\Gamma^+_4$ & $Y^+_4$ & (a) & (a) & 1 & -1 & -1 & 1 & 1 & -1 & -1 & 1\\
$\Gamma^-_1$ & $Y^-_1$ & (a) & (a) & 1 & 1 & 1 & 1 & -1 & -1 & -1 & -1\\
$\Gamma^-_2$ & $Y^-_2$ & (a) & (a) & 1 & -1 & 1 & -1 & -1 & 1 & -1 & 1\\
$\Gamma^-_3$ & $Y^-_3$ & (a) & (a) & 1 & 1 & -1 & -1 & -1 & -1 & 1 & 1\\
$\Gamma^-_4$ & $Y^-_4$ & (a) & (a) & 1 & -1 & -1 & 1 & -1 & 1 & 1 & -1\\
\hline\\
\end{tabular}
\begin{tabular}[t]{ccccccccccccc}
\multicolumn{13}{c}{$Z(00\frac{1}{2})$ and $T(\overline{\frac{1}{2}}\frac{1}{2}\frac{1}{2})$}\\
 &  &  & $$ & $$ & $\{C_{2a}|\frac{1}{2}\frac{1}{2}\frac{3}{2}\}$ & $\{I|001\}$ & $\{\sigma_{da}|\frac{1}{2}\frac{1}{2}\frac{3}{2}\}$ & $$ & $$ & $\{\sigma_z|\frac{1}{2}\frac{1}{2}\frac{3}{2}\}$ & $\{C_{2b}|001\}$ & $\{C_{2z}|\frac{1}{2}\frac{1}{2}\frac{3}{2}\}$\\
 & $K$ & $\{K|\frac{1}{2}\frac{1}{2}\frac{1}{2}\}$ & $\{E|000\}$ & $\{E|001\}$ & $\{C_{2a}|\frac{1}{2}\frac{1}{2}\frac{1}{2}\}$ & $\{I|000\}$ & $\{\sigma_{da}|\frac{1}{2}\frac{1}{2}\frac{1}{2}\}$ & $\{\sigma_{db}|000\}$ & $\{\sigma_{db}|001\}$ & $\{\sigma_z|\frac{1}{2}\frac{1}{2}\frac{1}{2}\}$ & $\{C_{2b}|000\}$ & $\{C_{2z}|\frac{1}{2}\frac{1}{2}\frac{1}{2}\}$\\
\hline
$Z_1$ & (a) & (a) & 2 & -2 & 0 & 0 & 0 & 2 & -2 & 0 & 0 & 0\\
$Z_2$ & (a) & (a) & 2 & -2 & 0 & 0 & 0 & -2 & 2 & 0 & 0 & 0\\
\hline\\
\end{tabular}
\begin{tabular}[t]{cccccccc}
\multicolumn{8}{c}{$S(0\frac{1}{2}0)$}\\
 &  &  & $$ & $$ & $\{\sigma_{da}|\frac{1}{2}\frac{3}{2}\frac{1}{2}\}$ & $\{I|010\}$ & $\{C_{2a}|\frac{1}{2}\frac{3}{2}\frac{1}{2}\}$\\
 & $K$ & $\{K|\frac{1}{2}\frac{1}{2}\frac{1}{2}\}$ & $\{E|000\}$ & $\{E|010\}$ & $\{\sigma_{da}|\frac{1}{2}\frac{1}{2}\frac{1}{2}\}$ & $\{I|000\}$ & $\{C_{2a}|\frac{1}{2}\frac{1}{2}\frac{1}{2}\}$\\
\hline
$S_1$ & (a) & (a) & 2 & -2 & 0 & 0 & 0\\
\hline\\
\end{tabular}
\begin{tabular}[t]{ccccccccccc}
\multicolumn{11}{c}{$R(0\frac{1}{2}\frac{1}{2})$}\\
 & $K$ & $\{K|\frac{1}{2}\frac{1}{2}\frac{1}{2}\}$ & $\{E|000\}$ & $\{\sigma_{da}|\frac{1}{2}\frac{1}{2}\frac{1}{2}\}$ & $\{E|001\}$ & $\{\sigma_{da}|\frac{1}{2}\frac{1}{2}\frac{3}{2}\}$ & $\{I|000\}$ & $\{C_{2a}|\frac{1}{2}\frac{1}{2}\frac{1}{2}\}$ & $\{I|001\}$ & $\{C_{2a}|\frac{1}{2}\frac{1}{2}\frac{3}{2}\}$\\
\hline
$R^+_1$ & (c) & (a) & 1 & i & -1 & -i & 1 & i & -1 & -i\\
$R^+_2$ & (c) & (a) & 1 & -i & -1 & i & 1 & -i & -1 & i\\
$R^-_1$ & (c) & (a) & 1 & i & -1 & -i & -1 & -i & 1 & i\\
$R^-_2$ & (c) & (a) & 1 & -i & -1 & i & -1 & i & 1 & -i\\
\hline\\
\end{tabular}
\end{table*}

\begin{table*}[H]
\caption{
  Characters of the small (allowed) single-valued irreducible
  representations of the space 
  group $H = Pccn = D^{10}_{2h}$ (56) of the antiferromagnetic structure of
  La$_2$CuO$_4$ in the notation of Table 5.8 in the textbook of Bradley and
  Cracknell \protect\cite{bc}.  $K$ stands for the operator of 
  time-inversion. The 
  first three elements in the first row define three magnetic groups $M_1 =
  H + KH, M_2 = H + \{K|\frac{1}{2}\frac{1}{2}0\}H,\mbox{ and } M_3 = H +
  \{K|\frac{1}{2}0\frac{1}{2}\}H$. The second, third, and fourth column of
  the table specify whether the corepresentations of $M_1$, $M_2$, and
  $M_3$, respectively, derived from the given representations $R_i$ of $H$
  follow case $(a)$, case $(b)$, or case $(c)$ when they are given by Eqs.
  (7.3.45-47) of Ref.~\protect\cite{bc}. In case (a), there is no change in
  the degeneracy of $R_i$, and in case (c) the degeneracy becomes doubled.
\label{tab:edaf}
}
\begin{tabular}[t]{cccccccccccc}
\multicolumn{12}{c}{$\Gamma (000)$}\\
 & $K$ & $\{K|\frac{1}{2}\frac{1}{2}0\}$ & $\{K|\frac{1}{2}0\frac{1}{2}\}$ & $\{E|000\}$ & $\{C_{2z}|\frac{1}{2}\frac{1}{2}0\}$ & $\{C_{2b}|\frac{1}{2}0\frac{1}{2}\}$ & $\{C_{2a}|0\frac{1}{2}\frac{1}{2}\}$ & $\{I|000\}$ & $\{\sigma_z|\frac{1}{2}\frac{1}{2}0\}$ & $\{\sigma_{db}|\frac{1}{2}0\frac{1}{2}\}$ & $\{\sigma_{da}|0\frac{1}{2}\frac{1}{2}\}$\\
\hline
$\Gamma^+_1$ & (a) & (a) & (a) & 1 & 1 & 1 & 1 & 1 & 1 & 1 & 1\\
$\Gamma^+_2$ & (a) & (a) & (a) & 1 & -1 & 1 & -1 & 1 & -1 & 1 & -1\\
$\Gamma^+_3$ & (a) & (a) & (a) & 1 & 1 & -1 & -1 & 1 & 1 & -1 & -1\\
$\Gamma^+_4$ & (a) & (a) & (a) & 1 & -1 & -1 & 1 & 1 & -1 & -1 & 1\\
$\Gamma^-_1$ & (a) & (a) & (a) & 1 & 1 & 1 & 1 & -1 & -1 & -1 & -1\\
$\Gamma^-_2$ & (a) & (a) & (a) & 1 & -1 & 1 & -1 & -1 & 1 & -1 & 1\\
$\Gamma^-_3$ & (a) & (a) & (a) & 1 & 1 & -1 & -1 & -1 & -1 & 1 & 1\\
$\Gamma^-_4$ & (a) & (a) & (a) & 1 & -1 & -1 & 1 & -1 & 1 & 1 & -1\\
\hline\\
\end{tabular}
\begin{tabular}[t]{cccccccc}
\multicolumn{8}{c}{$Y (\overline{\frac{1}{2}}00)$}\\
 &  &  &  & $$ & $$ & $\{\sigma_z|\frac{3}{2}\frac{1}{2}0\}$ & $\{I|000\}$\\
 & $K$ & $\{K|\frac{1}{2}\frac{1}{2}0\}$ & $\{K|\frac{1}{2}0\frac{1}{2}\}$ & $\{E|000\}$ & $\{E|100\}$ & $\{\sigma_z|\frac{1}{2}\frac{1}{2}0\}$ & $\{I|100\}$\\
\hline
$Y_1$ & (a) & (a) & (a) & 2 & -2 & 0 & 0\\
$Y_2$ & (a) & (a) & (a) & 2 & -2 & 0 & 0\\
\hline\\
\end{tabular}
\begin{tabular}[t]{ccccccc}
\multicolumn{7}{c}{$Y (\overline{\frac{1}{2}}00)$\qquad $(continued)$}\\
 & $\{C_{2z}|\frac{1}{2}\frac{1}{2}0\}$ & $$ & $$ & $\{C_{2b}|\frac{3}{2}0\frac{1}{2}\}$ & $\{C_{2a}|0\frac{1}{2}\frac{1}{2}\}$ & $\{\sigma_{db}|\frac{1}{2}0\frac{1}{2}\}$\\
 & $\{C_{2z}|\frac{3}{2}\frac{1}{2}0\}$ & $\{\sigma_{da}|0\frac{1}{2}\frac{1}{2}\}$ & $\{\sigma_{da}|1\frac{1}{2}\frac{1}{2}\}$ & $\{C_{2b}|\frac{1}{2}0\frac{1}{2}\}$ & $\{C_{2a}|1\frac{1}{2}\frac{1}{2}\}$ & $\{\sigma_{db}|\frac{3}{2}0\frac{1}{2}\}$\\
\hline
$Y_1$ & 0 & 2 & -2 & 0 & 0 & 0\\
$Y_2$ & 0 & -2 & 2 & 0 & 0 & 0\\
\hline\\
\end{tabular}
\begin{tabular}[t]{cccccccc}
\multicolumn{8}{c}{$X (0\frac{1}{2}0)$}\\
 &  &  &  & $$ & $$ & $\{\sigma_z|\frac{1}{2}\frac{3}{2}0\}$ & $\{I|010\}$\\
 & $K$ & $\{K|\frac{1}{2}\frac{1}{2}0\}$ & $\{K|\frac{1}{2}0\frac{1}{2}\}$ & $\{E|000\}$ & $\{E|010\}$ & $\{\sigma_z|\frac{1}{2}\frac{1}{2}0\}$ & $\{I|000\}$\\
\hline
$X_1$ & (a) & (a) & (a) & 2 & -2 & 0 & 0\\
$X_2$ & (a) & (a) & (a) & 2 & -2 & 0 & 0\\
\hline\\
\end{tabular}
\begin{tabular}[t]{ccccccc}
\multicolumn{7}{c}{$X (0\frac{1}{2}0)$\qquad $(continued)$}\\
 & $\{C_{2z}|\frac{1}{2}\frac{3}{2}0\}$ & $$ & $$ & $\{C_{2a}|0\frac{3}{2}\frac{1}{2}\}$ & $\{C_{2b}|\frac{1}{2}1\frac{1}{2}\}$ & $\{\sigma_{da}|0\frac{3}{2}\frac{1}{2}\}$\\
 & $\{C_{2z}|\frac{1}{2}\frac{1}{2}0\}$ & $\{\sigma_{db}|\frac{1}{2}0\frac{1}{2}\}$ & $\{\sigma_{db}|\frac{1}{2}1\frac{1}{2}\}$ & $\{C_{2a}|0\frac{1}{2}\frac{1}{2}\}$ & $\{C_{2b}|\frac{1}{2}0\frac{1}{2}\}$ & $\{\sigma_{da}|0\frac{1}{2}\frac{1}{2}\}$\\
\hline
$X_1$ & 0 & 2 & -2 & 0 & 0 & 0\\
$X_2$ & 0 & -2 & 2 & 0 & 0 & 0\\
\hline\\
\end{tabular}
\begin{tabular}[t]{cccccccc}
\multicolumn{8}{c}{$Z (00\frac{1}{2})$}\\
 &  &  &  & $$ & $$ & $\{\sigma_{da}|0\frac{1}{2}\frac{3}{2}\}$ & $\{I|001\}$\\
 & $K$ & $\{K|\frac{1}{2}\frac{1}{2}0\}$ & $\{K|\frac{1}{2}0\frac{1}{2}\}$ & $\{E|000\}$ & $\{E|001\}$ & $\{\sigma_{da}|0\frac{1}{2}\frac{1}{2}\}$ & $\{I|000\}$\\
\hline
$Z_1$ & (a) & (a) & (a) & 2 & -2 & 0 & 0\\
$Z_2$ & (a) & (a) & (a) & 2 & -2 & 0 & 0\\
\hline\\
\end{tabular}
\begin{tabular}[t]{ccccccc}
\multicolumn{7}{c}{$Z (00\frac{1}{2})$\qquad $(continued)$}\\
 & $\{C_{2a}|0\frac{1}{2}\frac{3}{2}\}$ & $$ & $$ & $\{\sigma_{db}|\frac{1}{2}0\frac{3}{2}\}$ & $\{\sigma_z|\frac{1}{2}\frac{1}{2}1\}$ & $\{C_{2b}|\frac{1}{2}0\frac{3}{2}\}$\\
 & $\{C_{2a}|0\frac{1}{2}\frac{1}{2}\}$ & $\{C_{2z}|\frac{1}{2}\frac{1}{2}0\}$ & $\{C_{2z}|\frac{1}{2}\frac{1}{2}1\}$ & $\{\sigma_{db}|\frac{1}{2}0\frac{1}{2}\}$ & $\{\sigma_z|\frac{1}{2}\frac{1}{2}0\}$ & $\{C_{2b}|\frac{1}{2}0\frac{1}{2}\}$\\
\hline
$Z_1$ & 0 & 2 & -2 & 0 & 0 & 0\\
$Z_2$ & 0 & -2 & 2 & 0 & 0 & 0\\
\hline\\
\end{tabular}
\end{table*}
\addtocounter{table}{-1}
\begin{table*}
\caption{$continued$}
\begin{tabular}[t]{cccccccc}
\multicolumn{8}{c}{$U (0\frac{1}{2}\frac{1}{2})$}\\
 & $K$ & $\{K|\frac{1}{2}\frac{1}{2}0\}$ & $\{K|\frac{1}{2}0\frac{1}{2}\}$ & $\{E|000\}$ & $\{\sigma_{da}|0\frac{1}{2}\frac{3}{2}\}$ & $\{E|001\}$ & $\{\sigma_{da}|0\frac{1}{2}\frac{1}{2}\}$\\
\hline
$U_1$ & (c) & (a) & (c) & 2 & 2i & -2 & -2i\\
$U_2$ & (c) & (a) & (c) & 2 & -2i & -2 & 2i\\
\hline\\
\end{tabular}
\begin{tabular}[t]{ccccccc}
\multicolumn{7}{c}{$U (0\frac{1}{2}\frac{1}{2})$\qquad $(continued)$}\\
 & $\{C_{2z}|\frac{1}{2}\frac{1}{2}0\}$ & $\{\sigma_{db}|\frac{1}{2}0\frac{3}{2}\}$ & $\{I|001\}$ & $\{C_{2a}|0\frac{1}{2}\frac{1}{2}\}$ & $\{\sigma_z|\frac{1}{2}\frac{1}{2}0\}$ & $\{C_{2b}|\frac{1}{2}0\frac{3}{2}\}$\\
 & $\{C_{2z}|\frac{1}{2}\frac{1}{2}1\}$ & $\{\sigma_{db}|\frac{1}{2}0\frac{1}{2}\}$ & $\{I|000\}$ & $\{C_{2a}|0\frac{1}{2}\frac{3}{2}\}$ & $\{\sigma_z|\frac{1}{2}\frac{1}{2}1\}$ & $\{C_{2b}|\frac{1}{2}0\frac{1}{2}\}$\\
\hline
$U_1$ & 0 & 0 & 0 & 0 & 0 & 0\\
$U_2$ & 0 & 0 & 0 & 0 & 0 & 0\\
\hline\\
\end{tabular}
\begin{tabular}[t]{cccccccc}
\multicolumn{8}{c}{$T (\overline{\frac{1}{2}}0\frac{1}{2})$}\\
 & $K$ & $\{K|\frac{1}{2}\frac{1}{2}0\}$ & $\{K|\frac{1}{2}0\frac{1}{2}\}$ & $\{E|000\}$ & $\{\sigma_{db}|\frac{1}{2}0\frac{1}{2}\}$ & $\{E|001\}$ & $\{\sigma_{db}|\frac{1}{2}0\frac{3}{2}\}$\\
\hline
$T_1$ & (c) & (a) & (a) & 2 & 2i & -2 & -2i\\
$T_2$ & (c) & (a) & (a) & 2 & -2i & -2 & 2i\\
\hline\\
\end{tabular}
\begin{tabular}[t]{ccccccc}
\multicolumn{7}{c}{$T (\overline{\frac{1}{2}}0\frac{1}{2})$\qquad $(continued)$}\\
 & $\{C_{2z}|\frac{1}{2}\frac{1}{2}1\}$ & $\{\sigma_{da}|0\frac{1}{2}\frac{3}{2}\}$ & $\{I|000\}$ & $\{C_{2b}|\frac{1}{2}0\frac{1}{2}\}$ & $\{\sigma_z|\frac{1}{2}\frac{1}{2}0\}$ & $\{C_{2a}|0\frac{1}{2}\frac{1}{2}\}$\\
 & $\{C_{2z}|\frac{1}{2}\frac{1}{2}0\}$ & $\{\sigma_{da}|0\frac{1}{2}\frac{1}{2}\}$ & $\{I|001\}$ & $\{C_{2b}|\frac{1}{2}0\frac{3}{2}\}$ & $\{\sigma_z|\frac{1}{2}\frac{1}{2}1\}$ & $\{C_{2a}|0\frac{1}{2}\frac{3}{2}\}$\\
\hline
$T_1$ & 0 & 0 & 0 & 0 & 0 & 0\\
$T_2$ & 0 & 0 & 0 & 0 & 0 & 0\\
\hline\\
\end{tabular}
\begin{tabular}[t]{cccccccc}
\multicolumn{8}{c}{$S (\overline{\frac{1}{2}}\frac{1}{2}0)$}\\
 &  &  &  & $$ & $$ & $\{C_{2b}|\frac{1}{2}1\frac{1}{2}\}$ & $\{I|000\}$\\
 & $K$ & $\{K|\frac{1}{2}\frac{1}{2}0\}$ & $\{K|\frac{1}{2}0\frac{1}{2}\}$ & $\{E|000\}$ & $\{E|010\}$ & $\{C_{2b}|\frac{1}{2}0\frac{1}{2}\}$ & $\{I|010\}$\\
\hline
$S_1$ & (a) & (a) & (c) & 2 & -2 & 0 & 0\\
$S_2$ & (a) & (a) & (c) & 2 & -2 & 0 & 0\\
\hline\\
\end{tabular}
\begin{tabular}[t]{ccccccc}
\multicolumn{7}{c}{$S (\overline{\frac{1}{2}}\frac{1}{2}0)$\qquad $(continued)$}\\
 & $\{\sigma_{db}|\frac{1}{2}0\frac{1}{2}\}$ & $$ & $$ & $\{C_{2a}|0\frac{1}{2}\frac{1}{2}\}$ & $\{\sigma_z|\frac{1}{2}\frac{3}{2}0\}$ & $\{\sigma_{da}|0\frac{3}{2}\frac{1}{2}\}$\\
 & $\{\sigma_{db}|\frac{1}{2}1\frac{1}{2}\}$ & $\{C_{2z}|\frac{1}{2}\frac{3}{2}0\}$ & $\{C_{2z}|\frac{1}{2}\frac{1}{2}0\}$ & $\{C_{2a}|0\frac{3}{2}\frac{1}{2}\}$ & $\{\sigma_z|\frac{1}{2}\frac{1}{2}0\}$ & $\{\sigma_{da}|0\frac{1}{2}\frac{1}{2}\}$\\
\hline
$S_1$ & 0 & 2 & -2 & 0 & 0 & 0\\
$S_2$ & 0 & -2 & 2 & 0 & 0 & 0\\
\hline\\
\end{tabular}
\begin{tabular}[t]{ccccccccccc}
\multicolumn{11}{c}{$R (\overline{\frac{1}{2}}\frac{1}{2}\frac{1}{2})$}\\
 & $K$ & $\{K|\frac{1}{2}\frac{1}{2}0\}$ & $\{K|\frac{1}{2}0\frac{1}{2}\}$ & $\{E|000\}$ & $\{C_{2a}|0\frac{1}{2}\frac{1}{2}\}$ & $\{E|001\}$ & $\{C_{2a}|0\frac{1}{2}\frac{3}{2}\}$ & $\{C_{2z}|\frac{1}{2}\frac{1}{2}1\}$ & $\{C_{2b}|\frac{1}{2}0\frac{3}{2}\}$ & $\{C_{2z}|\frac{1}{2}\frac{1}{2}0\}$\\
\hline
$R^-_1$ & (c) & (a) & (c) & 1 & i & -1 & -i & 1 & i & -1\\
$R^-_2$ & (c) & (a) & (c) & 1 & -i & -1 & i & 1 & -i & -1\\
$R^-_3$ & (c) & (a) & (c) & 1 & i & -1 & -i & -1 & -i & 1\\
$R^-_4$ & (c) & (a) & (c) & 1 & -i & -1 & i & -1 & i & 1\\
$R^+_1$ & (c) & (a) & (c) & 1 & i & -1 & -i & 1 & i & -1\\
$R^+_2$ & (c) & (a) & (c) & 1 & -i & -1 & i & 1 & -i & -1\\
$R^+_3$ & (c) & (a) & (c) & 1 & i & -1 & -i & -1 & -i & 1\\
$R^+_4$ & (c) & (a) & (c) & 1 & -i & -1 & i & -1 & i & 1\\
\hline\\
\end{tabular}
\begin{tabular}[t]{cccccccccc}
\multicolumn{10}{c}{$R (\overline{\frac{1}{2}}\frac{1}{2}\frac{1}{2})$\qquad $(continued)$}\\
 & $\{C_{2b}|\frac{1}{2}0\frac{1}{2}\}$ & $\{I|001\}$ & $\{\sigma_{da}|0\frac{1}{2}\frac{3}{2}\}$ & $\{I|000\}$ & $\{\sigma_{da}|0\frac{1}{2}\frac{1}{2}\}$ & $\{\sigma_z|\frac{1}{2}\frac{1}{2}0\}$ & $\{\sigma_{db}|\frac{1}{2}0\frac{1}{2}\}$ & $\{\sigma_z|\frac{1}{2}\frac{1}{2}1\}$ & $\{\sigma_{db}|\frac{1}{2}0\frac{3}{2}\}$\\
\hline
$R^-_1$ & -i & 1 & i & -1 & -i & 1 & i & -1 & -i\\
$R^-_2$ & i & 1 & -i & -1 & i & 1 & -i & -1 & i\\
$R^-_3$ & i & 1 & i & -1 & -i & -1 & -i & 1 & i\\
$R^-_4$ & -i & 1 & -i & -1 & i & -1 & i & 1 & -i\\
$R^+_1$ & -i & -1 & -i & 1 & i & -1 & -i & 1 & i\\
$R^+_2$ & i & -1 & i & 1 & -i & -1 & i & 1 & -i\\
$R^+_3$ & i & -1 & -i & 1 & i & 1 & i & -1 & -i\\
$R^+_4$ & -i & -1 & i & 1 & -i & 1 & -i & -1 & i\\
\hline\\
\end{tabular}
\end{table*}

\begin{table*}[H]
\caption{
  Characters of the small (allowed) double-valued irreducible
  representations of the space 
  group $I4/mmm$ of tetragonal La$_2$CuO$_4$ in the notation of Table 5.8 in
  the textbook of Bradley and Cracknell \protect\cite{bc}.
\label{tab:zdtetra}
}
\begin{tabular}[t]{cccccccccccccccc}
\multicolumn{16}{c}{$\Gamma(000)$ and $Z(\frac{1}{2}\frac{1}{2}\overline{\frac{1}{2}})$}\\
& & $$ & $$ & $$ & $$ & $$ & $\overline{C}_{2x}$ & $\overline{C}_{2b}$ & $$ & $$ & $$ & $$ & $$ & $\overline{\sigma}_x$ & $\overline{\sigma}_{db}$\\
& & $$ & $$ & $$ & $$ & $$ & $\overline{C}_{2y}$ & $C_{2a}$ & $$ & $$ & $$ & $$ & $$ & $\overline{\sigma}_y$ & $\sigma_{da}$\\
& & $$ & $$ & $C_{2z}$ & $C^-_{4z}$ & $\overline{C}^+_{4z}$ & $C_{2y}$ & $C_{2b}$ & $$ & $$ & $\sigma_z$ & $S^+_{4z}$ & $\overline{S}^-_{4z}$ & $\sigma_y$ & $\sigma_{db}$\\
& & $E$ & $\overline{E}$ & $\overline{C}_{2z}$ & $C^+_{4z}$ & $\overline{C}^-_{4z}$ & $C_{2x}$ & $\overline{C}_{2a}$ & $I$ & $\overline{I}$ & $\overline{\sigma}_z$ & $S^-_{4z}$ & $\overline{S}^+_{4z}$ & $\sigma_x$ & $\overline{\sigma}_{da}$\\
\hline
$\Gamma^+_6$ & $Z^+_6$ & 2 & -2 & 0 & $\sqrt{2}$ & -$\sqrt{2}$ & 0 & 0 & 2 & -2 & 0 & $\sqrt{2}$ & -$\sqrt{2}$ & 0 & 0\\
$\Gamma^+_7$ & $Z^+_7$ & 2 & -2 & 0 & -$\sqrt{2}$ & $\sqrt{2}$ & 0 & 0 & 2 & -2 & 0 & -$\sqrt{2}$ & $\sqrt{2}$ & 0 & 0\\
$\Gamma^-_6$ & $Z^-_6$ & 2 & -2 & 0 & $\sqrt{2}$ & -$\sqrt{2}$ & 0 & 0 & -2 & 2 & 0 & -$\sqrt{2}$ & $\sqrt{2}$ & 0 & 0\\
$\Gamma^-_7$ & $Z^-_7$ & 2 & -2 & 0 & -$\sqrt{2}$ & $\sqrt{2}$ & 0 & 0 & -2 & 2 & 0 & $\sqrt{2}$ & -$\sqrt{2}$ & 0 & 0\\
\hline\\
\end{tabular}\\
\begin{tabular}[t]{ccccccccc}
\multicolumn{9}{c}{$N(0\frac{1}{2}0)$}\\
 & $E$ & $C_{2y}$ & $\overline{E}$ & $\overline{C}_{2y}$ & $I$ & $\sigma_y$ & $\overline{I}$ & $\overline{\sigma}_y$\\
\hline
$N^+_3$ & 1 & i & -1 & -i & 1 & i & -1 & -i\\
$N^+_4$ & 1 & -i & -1 & i & 1 & -i & -1 & i\\
$N^-_3$ & 1 & i & -1 & -i & -1 & -i & 1 & i\\
$N^-_4$ & 1 & -i & -1 & i & -1 & i & 1 & -i\\
\hline\\
\end{tabular}\\

\begin{tabular}[t]{ccccccccccc}
\multicolumn{11}{c}{$X(00\frac{1}{2})$}\\
 & $$ & $$ & $\overline{C}_{2z}$ & $\overline{C}_{2a}$ & $C_{2b}$ & $$ & $$ & $\overline{\sigma}_z$ & $\overline{\sigma}_{da}$ & $\sigma_{db}$\\
 & $E$ & $\overline{E}$ & $C_{2z}$ & $C_{2a}$ & $\overline{C}_{2b}$ & $I$ & $\overline{I}$ & $\sigma_z$ & $\sigma_{da}$ & $\overline{\sigma}_{db}$\\
\hline
$X^+_5$ & 2 & -2 & 0 & 0 & 0 & 2 & -2 & 0 & 0 & 0\\
$X^-_5$ & 2 & -2 & 0 & 0 & 0 & -2 & 2 & 0 & 0 & 0\\
\hline\\
\end{tabular}\\
\begin{tabular}[t]{cccccccc}
\multicolumn{8}{c}{$P(\frac{1}{4}\frac{1}{4}\frac{1}{4})$}\\
 & $$ & $$ & $$ & $$ & $$ & $\overline{C}_{2x}$ & $\sigma_{da}$\\
 & $$ & $$ & $$ & $$ & $$ & $C_{2y}$ & $\overline{\sigma}_{db}$\\
 & $$ & $$ & $\overline{C}_{2z}$ & $S^-_{4z}$ & $\overline{S}^+_{4z}$ & $\overline{C}_{2y}$ & $\overline{\sigma}_{da}$\\
 & $E$ & $\overline{E}$ & $C_{2z}$ & $S^+_{4z}$ & $\overline{S}^-_{4z}$ & $C_{2x}$ & $\sigma_{db}$\\
\hline
$P_6$ & 2 & -2 & 0 & $\sqrt{2}$ & -$\sqrt{2}$ & 0 & 0\\
$P_7$ & 2 & -2 & 0 & -$\sqrt{2}$ & $\sqrt{2}$ & 0 & 0\\
\hline\\
\end{tabular}
\end{table*}

\begin{table*}[H]
\caption{
  Characters of the small (allowed) double-valued irreducible
  representations of the space 
  group $Cmca$ of the orthorhombic phase of La$_2$CuO$_4$ in the notation of
  Table 5.8 in the textbook of Bradley and Cracknell \protect\cite{bc}.
\label{tab:zdortho}
}
\begin{tabular}[t]{cccccccccccc}
\multicolumn{12}{c}{$\Gamma(000)$ and
 $Y(\overline{\frac{1}{2}}\frac{1}{2}0)$}\\ 
& & $$ & $$ & $\{\overline{C}_{2a}|\frac{1}{2}\frac{1}{2}\frac{1}{2}\}$ & $\{\overline{C}_{2z}|\frac{1}{2}\frac{1}{2}\frac{1}{2}\}$ & $\{\overline{C}_{2b}|000\}$ & $$ & $$ & $\{\overline{\sigma}_{da}|\frac{1}{2}\frac{1}{2}\frac{1}{2}\}$ & $\{\overline{\sigma}_z|\frac{1}{2}\frac{1}{2}\frac{1}{2}\}$ & $\{\overline{\sigma}_{db}|000\}$\\
& & $\{E|000\}$ & $\{\overline{E}|000\}$ & $\{C_{2a}|\frac{1}{2}\frac{1}{2}\frac{1}{2}\}$ & $\{C_{2z}|\frac{1}{2}\frac{1}{2}\frac{1}{2}\}$ & $\{C_{2b}|000\}$ & $\{I|000\}$ & $\{\overline{I}|000\}$ & $\{\sigma_{da}|\frac{1}{2}\frac{1}{2}\frac{1}{2}\}$ & $\{\sigma_z|\frac{1}{2}\frac{1}{2}\frac{1}{2}\}$ & $\{\sigma_{db}|000\}$\\
\hline
$\Gamma^+_5$ & $Y^+_5$ & 2 & -2 & 0 & 0 & 0 & 2 & -2 & 0 & 0 & 0\\
$\Gamma^-_5$ & $Y^-_5$ &2 & -2 & 0 & 0 & 0 & -2 & 2 & 0 & 0 & 0\\
\hline\\
\end{tabular}
\begin{tabular}{cccccccccc}
\multicolumn{9}{c}{$Z(00\frac{1}{2})$ and
 $T(\overline{\frac{1}{2}}\frac{1}{2}\frac{1}{2})$}\\ 
& & $$ & $$ & $$ & $$ & $\{\overline{C}_{2a}|\frac{1}{2}\frac{1}{2}\frac{1}{2}\}$ & $$ & $$\\
& & $$ & $$ & $$ & $$ & $\{\overline{C}_{2a}|\frac{1}{2}\frac{1}{2}\frac{3}{2}\}$ & $$ & $$\\
& & $$ & $$ & $$ & $$ & $\{C_{2a}|\frac{1}{2}\frac{1}{2}\frac{3}{2}\}$ & $\{\overline{C}_{2b}|001\}$ & $\{C_{2b}|001\}$\\
& & $\{E|000\}$ & $\{\overline{E}|001\}$ & $\{\overline{E}|000\}$ & $\{E|001\}$ & $\{C_{2a}|\frac{1}{2}\frac{1}{2}\frac{1}{2}\}$ & $\{C_{2b}|000\}$ & $\{\overline{C}_{2b}|000\}$\\
\hline
$Z_3$ & $T_3$ & 2 & 2 & -2 & -2 & 0 & 2i & -2i\\
$Z_4$ & $T_4$ & 2 & 2 & -2 & -2 & 0 & -2i & 2i\\
\hline\\
\end{tabular}
\begin{tabular}{ccccccccc}
\multicolumn{9}{c}{$Z(00\frac{1}{2})$ and
 $T(\overline{\frac{1}{2}}\frac{1}{2}\frac{1}{2})$\qquad $(continued)$}\\ 
& & $\{C_{2z}|\frac{1}{2}\frac{1}{2}\frac{1}{2}\}$ & $$ & $$ & $\{\overline{\sigma}_{da}|\frac{1}{2}\frac{1}{2}\frac{1}{2}\}$ & $$ & $$ & $\{\sigma_z|\frac{1}{2}\frac{1}{2}\frac{1}{2}\}$\\
& & $\{C_{2z}|\frac{1}{2}\frac{1}{2}\frac{3}{2}\}$ & $$ & $$ & $\{\overline{\sigma}_{da}|\frac{1}{2}\frac{1}{2}\frac{3}{2}\}$ & $$ & $$ & $\{\sigma_z|\frac{1}{2}\frac{1}{2}\frac{3}{2}\}$\\
& & $\{\overline{C}_{2z}|\frac{1}{2}\frac{1}{2}\frac{3}{2}\}$ & $\{I|001\}$ & $\{\overline{I}|000\}$ & $\{\sigma_{da}|\frac{1}{2}\frac{1}{2}\frac{3}{2}\}$ & $\{\overline{\sigma}_{db}|000\}$ & $\{\sigma_{db}|001\}$ & $\{\overline{\sigma}_z|\frac{1}{2}\frac{1}{2}\frac{3}{2}\}$\\
& & $\{\overline{C}_{2z}|\frac{1}{2}\frac{1}{2}\frac{1}{2}\}$ & $\{I|000\}$ & $\{\overline{I}|001\}$ & $\{\sigma_{da}|\frac{1}{2}\frac{1}{2}\frac{1}{2}\}$ & $\{\sigma_{db}|000\}$ & $\{\overline{\sigma}_{db}|001\}$ & $\{\overline{\sigma}_z|\frac{1}{2}\frac{1}{2}\frac{1}{2}\}$\\
\hline
$Z_3$ & $T_3$ & 0 & 0 & 0 & 0 & 0 & 0 & 0\\
$Z_4$ & $T_4$ & 0 & 0 & 0 & 0 & 0 & 0 & 0\\
\hline\\
\end{tabular}
\begin{tabular}{ccccccccccc}
\multicolumn{11}{c}{$S(0\frac{1}{2}0)$}\\
 & $$ & $$ & $$ & $$ & $\{I|010\}$ & $\{\overline{I}|000\}$ & $\{\sigma_{da}|\frac{1}{2}\frac{3}{2}\frac{1}{2}\}$ & $\{\overline{\sigma}_{da}|\frac{1}{2}\frac{1}{2}\frac{1}{2}\}$ & $\{C_{2a}|\frac{1}{2}\frac{3}{2}\frac{1}{2}\}$ & $\{\overline{C}_{2a}|\frac{1}{2}\frac{1}{2}\frac{1}{2}\}$\\
 & $\{E|000\}$ & $\{E|010\}$ & $\{\overline{E}|010\}$ & $\{\overline{E}|000\}$ & $\{I|000\}$ & $\{\overline{I}|010\}$ & $\{\sigma_{da}|\frac{1}{2}\frac{1}{2}\frac{1}{2}\}$ & $\{\overline{\sigma}_{da}|\frac{1}{2}\frac{3}{2}\frac{1}{2}\}$ & $\{C_{2a}|\frac{1}{2}\frac{1}{2}\frac{1}{2}\}$ & $\{\overline{C}_{2a}|\frac{1}{2}\frac{3}{2}\frac{1}{2}\}$\\
\hline
$S_2$ & 2 & -2 & 2 & -2 & 0 & 0 & 0 & 0 & 0 & 0\\
\hline\\
\end{tabular}
\begin{tabular}{ccccccccc}
\multicolumn{9}{c}{$R(0\frac{1}{2}\frac{1}{2})$}\\
 & $\{E|000\}$ & $\{\sigma_{da}|\frac{1}{2}\frac{1}{2}\frac{1}{2}\}$ & $\{\overline{E}|001\}$ & $\{\overline{\sigma}_{da}|\frac{1}{2}\frac{1}{2}\frac{3}{2}\}$ & $\{I|000\}$ & $\{C_{2a}|\frac{1}{2}\frac{1}{2}\frac{1}{2}\}$ & $\{\overline{I}|001\}$ & $\{\overline{C}_{2a}|\frac{1}{2}\frac{1}{2}\frac{3}{2}\}$\\
\hline
$R^+_3$ & 1 & 1 & 1 & 1 & 1 & 1 & 1 & 1\\
$R^+_4$ & 1 & -1 & 1 & -1 & 1 & -1 & 1 & -1\\
$R^-_3$ & 1 & 1 & 1 & 1 & -1 & -1 & -1 & -1\\
$R^-_4$ & 1 & -1 & 1 & -1 & -1 & 1 & -1 & 1\\
\hline\\
\end{tabular}
\end{table*}

\begin{table*}
\caption{
Compatibility relations between the single-valued representations of the
space group $I4/mmm$ of tetragonal La$_2$CuO$_4$ and the single-valued
representations of the space group $Cmca$ of the orthorhombic phase of
La$_2$CuO$_4$. For the notations see Tables \protect\ref{tab:edtetra} and
\protect\ref{tab:edortho}. The upper row lists small
representations at the points of symmetry in the Brillouin zone of $I4/mmm$
and the lower row lists small representations at the related points of
symmetry in the Brillouin zone of $Cmca$ (cf. Fig.~\protect\ref{fig:bz}).
The small 
representations in the same column are compatible in the following sense:
Bloch functions which are basis functions of a small representation $R_i$ in
the upper row also form basis functions of the small representation standing
below $R_i$.
\label{tab:faltentetraortho}
}
\begin{tabular}[t]{cccccccccc}
\multicolumn{10}{c}{$\Gamma$}\\
\hline
$\Gamma^+_1$ & $\Gamma^+_2$ & $\Gamma^+_3$ & $\Gamma^+_4$ & $\Gamma^+_5$ & 
$\Gamma^-_1$ & $\Gamma^-_2$ & $\Gamma^-_3$ & $\Gamma^-_4$ & $\Gamma^-_5$\\
$\Gamma^+_1$ & $\Gamma^+_2$ & $\Gamma^+_2$ & $\Gamma^+_1$ & $\Gamma^+_3$
 + $\Gamma^+_4$ & $\Gamma^-_1$ & $\Gamma^-_2$ & $\Gamma^-_2$ & $\Gamma^-_1$ & 
$\Gamma^-_3$ + $\Gamma^-_4$\\
\hline\\
\end{tabular}\hspace{.7cm}
\begin{tabular}[t]{cccccccc}
\multicolumn{8}{c}{$\Gamma$}\\
\hline
$X^+_1$ & $X^+_2$ & $X^+_3$ & $X^+_4$ & $X^-_1$ & $X^-_2$ & $X^-_3$ & 
$X^-_4$\\
$\Gamma^+_4$ & $\Gamma^+_2$ & $\Gamma^+_3$ & $\Gamma^+_1$ & $\Gamma^-_4$ & 
$\Gamma^-_2$ & $\Gamma^-_3$ & $\Gamma^-_1$\\
\hline\\
\end{tabular}\hspace{.7cm}
\begin{tabular}[t]{cccccccccc}
\multicolumn{10}{c}{$Y$}\\
\hline
$Z^+_1$ & $Z^+_2$ & $Z^+_3$ & $Z^+_4$ & $Z^+_5$ & $Z^-_1$ & $Z^-_2$ & 
$Z^-_3$ & $Z^-_4$ & $Z^-_5$\\
$Y^+_1$ & $Y^+_2$ & $Y^+_2$ & $Y^+_1$ & $Y^+_3$ + $Y^+_4$ & $Y^-_1$ & 
$Y^-_2$ & $Y^-_2$ & $Y^-_1$ & $Y^-_3$ + $Y^-_4$\\
\hline\\
\end{tabular}\hspace{.7cm}
\begin{tabular}[t]{cccccccc}
\multicolumn{8}{c}{$Y$}\\
\hline
$X^+_1$ & $X^+_2$ & $X^+_3$ & $X^+_4$ & $X^-_1$ & $X^-_2$ & $X^-_3$ & 
$X^-_4$\\
$Y^+_4$ & $Y^+_1$ & $Y^+_3$ & $Y^+_2$ & $Y^-_4$ & $Y^-_1$ & $Y^-_3$ & 
$Y^-_2$\\
\hline\\
\end{tabular}\hspace{.7cm}
\begin{tabular}[t]{cccc}
\multicolumn{4}{c}{$R$}\\
\hline
$N^+_1$ & $N^-_1$ & $N^+_2$ & $N^-_2$\\
$R^+_1$ + $R^+_2$ & $R^-_1$ + $R^-_2$ & $R^+_1$ + $R^+_2$ & $R^-_1$ + $R^-_2$\\
\hline\\
\end{tabular}\hspace{.7cm}
\begin{tabular}[t]{cccc}
\multicolumn{4}{c}{$T$}\\
\hline
$U_1$ & $U_3$ & $U_2$ & $U_4$\\
$T_1$ & $T_2$ & $T_2$ & $T_1$\\
\hline\\
\end{tabular}\hspace{.7cm}
\begin{tabular}[t]{cccc}
\multicolumn{4}{c}{$Z$}\\
\hline
$\Delta_1$ & $\Delta_3$ & $\Delta_2$ & $\Delta_4$\\
$Z_1$ & $Z_2$ & $Z_2$ & $Z_1$\\
\hline\\
\end{tabular}\hspace{.7cm}
\begin{tabular}[t]{cc}
\multicolumn{2}{c}{$S$}\\
\hline
$E'_1$ & $E'_2$\\
$S_1$ & $S_1$\\
\hline\\
\end{tabular}
\end{table*}

\begin{table*}
\caption{
Compatibility relations between the single-valued representations of the
space group $I4/mmm$ of tetragonal La$_2$CuO$_4$ and the single-valued
representations of the space group $P4/mnc$ of antiferromagnetic
chromium. For the notations see Table \protect\ref{tab:edtetra} and Table II
of Ref.\ \protect\cite{eabf}. The upper row lists small
representations at the points of symmetry in the Brillouin zone of $I4/mmm$
and the lower row lists small representations at the related points of
symmetry in the Brillouin zone of $P4/mnc$.  The small
representations in the same column are compatible in the following sense:
Bloch functions which are basis functions of a small representation $R_i$ in
the upper row also form basis functions of the small representation standing
below $R_i$.
\label{tab:faltentetracraf}
}
\begin{tabular}[t]{cccccccccc}
\multicolumn{10}{c}{$\Gamma$}\\
\hline
$\Gamma^+_1$ & $\Gamma^+_2$ & $\Gamma^+_3$ & $\Gamma^+_4$ & $\Gamma^+_5$ & 
$\Gamma^-_1$ & $\Gamma^-_2$ & $\Gamma^-_3$ & $\Gamma^-_4$ & $\Gamma^-_5$\\
$\Gamma^+_1$ & $\Gamma^+_2$ & $\Gamma^+_3$ & $\Gamma^+_4$ & $\Gamma^+_5$ & 
$\Gamma^-_1$ & $\Gamma^-_2$ & $\Gamma^-_3$ & $\Gamma^-_4$ & $\Gamma^-_5$\\
\hline\\
\end{tabular}\hspace{.7cm}
\begin{tabular}[t]{cccccccccc}
\multicolumn{10}{c}{$\Gamma$}\\
\hline
$Z^+_1$ & $Z^+_2$ & $Z^+_3$ & $Z^+_4$ & $Z^+_5$ & $Z^-_1$ & $Z^-_2$ & 
$Z^-_3$ & $Z^-_4$ & $Z^-_5$\\
$\Gamma^+_2$ & $\Gamma^+_1$ & $\Gamma^+_4$ & $\Gamma^+_3$ & $\Gamma^+_5$ & 
$\Gamma^-_2$ & $\Gamma^-_1$ & $\Gamma^-_4$ & $\Gamma^-_3$ & $\Gamma^-_5$\\
\hline\\
\end{tabular}\hspace{.7cm}
\begin{tabular}[t]{cccccccc}
\multicolumn{8}{c}{$M$}\\
\hline
$X^+_1$ & $X^+_2$ & $X^+_3$ & $X^+_4$ & $X^-_1$ & $X^-_2$ & $X^-_3$ & 
$X^-_4$\\
$M_{20}$ & $M_{17}$ + $M_{18}$ & $M_{20}$ & $M_{15}$ + $M_{16}$ & $M_{10}$ & 
$M_{5}$ + $M_{6}$ & $M_{10}$ & $M_{7}$ + $M_{8}$\\
\hline\\
\end{tabular}\hspace{.7cm}
\begin{tabular}[t]{ccccc}
\multicolumn{5}{c}{$A$}\\
\hline
$P_1$ & $P_2$ & $P_3$ & $P_4$ & $P_5$\\
$A_{11}$ & $A_{11}$ & $A_{10}$ & $A_{10}$ & $A_{12}$ + $A_{13}$\\
\hline\\
\end{tabular}\\
\begin{tabular}[t]{cccc}
\multicolumn{4}{c}{$X$}\\
\hline
$\Sigma_1$ & $\Sigma_3$ & $\Sigma_2$ & $\Sigma_4$\\
$X_{5}$ & $X_{10}$ & $X_{5}$ & $X_{10}$\\
\hline\\
\end{tabular}\hspace{.7cm}
\begin{tabular}[t]{ccccc}
\multicolumn{5}{c}{$Z$}\\
\hline
$\Lambda_1$ & $\Lambda_2$ & $\Lambda_3$ & $\Lambda_4$ & $\Lambda_5$\\
$Z_{10}$ & $Z_{10}$ & $Z_{11}$ & $Z_{11}$ & $Z_{12}$ + $Z_{13}$\\
\hline\\
\end{tabular}\hspace{.7cm}
\begin{tabular}[t]{cccc}
\multicolumn{4}{c}{$R$}\\
\hline
$N^+_1$ & $N^-_1$ & $N^+_2$ & $N^-_2$\\
$R_{5}$ & $R_{10}$ & $R_{5}$ & $R_{10}$\\
\hline\\
\end{tabular}
\end{table*}

\begin{table*}
\caption{
Compatibility relations between the single-valued representations of the
space group $Cmca$ of the orthorhombic phase of La$_2$CuO$_4$ and the
single-valued representations of the space group $Pccn$ of the
antiferromagnetic structure of La$_2$CuO$_4$.  For the notations see Tables
\protect\ref{tab:edortho} and \protect\ref{tab:edaf}. The
upper row lists small representations at the points of symmetry in the
Brillouin zone of $Cmca$ and the lower row lists small representations at
the related points of symmetry in the Brillouin zone of $Pccn$ (cf.
Fig.~\protect\ref{fig:bz}).  The small representations in the same column
are compatible in the following sense: Bloch functions which are basis
functions of a small representation $R_i$ in the upper row also form basis
functions of the small representation standing below $R_i$.
\label{tab:faltenorthoaf}
}
\begin{tabular}[t]{cccccccc}
\multicolumn{8}{c}{$\Gamma$}\\
\hline
$\Gamma^+_1$ & $\Gamma^+_2$ & $\Gamma^+_3$ & $\Gamma^+_4$ & $\Gamma^-_1$ & 
$\Gamma^-_2$ & $\Gamma^-_3$ & $\Gamma^-_4$\\
$\Gamma^+_1$ & $\Gamma^+_3$ & $\Gamma^+_4$ & $\Gamma^+_2$ & $\Gamma^-_1$ & 
$\Gamma^-_3$ & $\Gamma^-_4$ & $\Gamma^-_2$\\
\hline\\
\end{tabular}\hspace{.7cm}
\begin{tabular}[t]{cccccccc}
\multicolumn{8}{c}{$\Gamma$}\\
\hline
$Y^+_1$ & $Y^+_2$ & $Y^+_3$ & $Y^+_4$ & $Y^-_1$ & $Y^-_2$ & $Y^-_3$ & 
$Y^-_4$\\
$\Gamma^+_3$ & $\Gamma^+_1$ & $\Gamma^+_2$ & $\Gamma^+_4$ & $\Gamma^-_3$ & 
$\Gamma^-_1$ & $\Gamma^-_2$ & $\Gamma^-_4$\\
\hline\\
\end{tabular}\\
\begin{tabular}[t]{cc}
\multicolumn{2}{c}{$X$}\\
\hline
$Z_1$ & $Z_2$\\
$X_1$ & $X_2$\\
\hline\\
\end{tabular}\hspace{.7cm}
\begin{tabular}[t]{cc}
\multicolumn{2}{c}{$X$}\\
\hline
$T_1$ & $T_2$\\
$X_2$ & $X_1$\\
\hline\\
\end{tabular}\hspace{.7cm}
\begin{tabular}[t]{c}
\multicolumn{1}{c}{$T$}\\
\hline
$S_1$\\
$T_1$ + $T_2$\\
\hline\\
\end{tabular}\hspace{.7cm}
\begin{tabular}[t]{c}
\multicolumn{1}{c}{$S$}\\
\hline
$A_1$\\
$S_1$ + $S_2$\\
\hline\\
\end{tabular}\hspace{.7cm}
\begin{tabular}[t]{cccc}
\multicolumn{4}{c}{$R$}\\
\hline
$R^+_1$ & $R^+_2$ & $R^-_1$ & $R^-_2$\\
$R^+_1$ + $R^+_3$ & $R^+_2$ + $R^+_4$ & $R^-_2$ + $R^-_4$ & $R^-_1$ + $R^-_3$\\
\hline\\
\end{tabular}\\
\begin{tabular}[t]{cccc}
\multicolumn{4}{c}{$Y$}\\
\hline
$\Sigma_1$ & $\Sigma_3$ & $\Sigma_2$ & $\Sigma_4$\\
$Y_1$ & $Y_2$ & $Y_1$ & $Y_2$\\
\hline\\
\end{tabular}\hspace{.7cm}
\begin{tabular}[t]{cccc}
\multicolumn{4}{c}{$U$}\\
\hline
$B_1$ & $B_2$ & $B_3$ & $B_4$\\
$U_1$ & $U_2$ & $U_2$ & $U_1$\\
\hline\\
\end{tabular}\hspace{.7cm}
\begin{tabular}[t]{cccc}
\multicolumn{4}{c}{$Z$}\\
\hline
$\Delta_1$ & $\Delta_3$ & $\Delta_2$ & $\Delta_4$\\
$Z_1$ & $Z_1$ & $Z_2$ & $Z_2$\\
\hline\\
\end{tabular}
\end{table*}
\begin{table*}
\caption{
  Compatibility relations between the single-valued (upper row) and
  double-valued (lower row) representations of the space group $I4/mmm$ of
  tetragonal La$_2$CuO$_4$. For the notations see Tables
  \protect\ref{tab:edtetra}
  and \protect\ref{tab:zdtetra}. The small representations in the same
  column are compatible in the following sense: spin-dependent Bloch
  functions which are basis functions of the
  small representation $R_i\times D_{1/2}$ (with $R_i$ standing in the upper
  row) also form basis functions of the double-valued small representation
  standing below $R_i$. 
  $D_{1/2}$ denotes the two-dimensional double-valued representation of the
  three-dimensional rotation group $O(3)$ (given, e.g., in Table 6.1 of
  Ref.~\protect\cite{bc}).
\label{tab:edzdtetra}
}
\begin{tabular}[t]{cccccccccc}
\multicolumn{10}{c}{  $\Gamma$}\\
\hline
$\Gamma^+_1$ & $\Gamma^+_2$ & $\Gamma^+_3$ & $\Gamma^+_4$ & $\Gamma^+_5$ & 
$\Gamma^-_1$ & $\Gamma^-_2$ & $\Gamma^-_3$ & $\Gamma^-_4$ & $\Gamma^-_5$\\
$\Gamma^+_6$ & $\Gamma^+_6$ & $\Gamma^+_7$ & $\Gamma^+_7$ & $\Gamma^+_6$
 + $\Gamma^+_7$ & $\Gamma^-_6$ & $\Gamma^-_6$ & $\Gamma^-_7$ & $\Gamma^-_7$ & 
$\Gamma^-_6$ + $\Gamma^-_7$\\
\hline\\
\end{tabular}\hspace{.7cm}
\begin{tabular}[t]{cccccccccc}
\multicolumn{10}{c}{  $Z$}\\
\hline
$Z^+_1$ & $Z^+_2$ & $Z^+_3$ & $Z^+_4$ & $Z^+_5$ & $Z^-_1$ & $Z^-_2$ & 
$Z^-_3$ & $Z^-_4$ & $Z^-_5$\\
$Z^+_6$ & $Z^+_6$ & $Z^+_7$ & $Z^+_7$ & $Z^+_6$ + $Z^+_7$ & $Z^-_6$ & 
$Z^-_6$ & $Z^-_7$ & $Z^-_7$ & $Z^-_6$ + $Z^-_7$\\
\hline\\
\end{tabular}\hspace{.7cm}
\begin{tabular}[t]{cccc}
\multicolumn{4}{c}{  $N$}\\
\hline
$N^+_1$ & $N^-_1$ & $N^+_2$ & $N^-_2$\\
$N^+_3$ + $N^+_4$ & $N^-_3$ + $N^-_4$ & $N^+_3$ + $N^+_4$ & $N^-_3$ + $N^-_4$\\
\hline\\
\end{tabular}\hspace{.7cm}
\begin{tabular}[t]{cccccccc}
\multicolumn{8}{c}{  $X$}\\
\hline
$X^+_1$ & $X^+_2$ & $X^+_3$ & $X^+_4$ & $X^-_1$ & $X^-_2$ & $X^-_3$ & 
$X^-_4$\\
$X^+_5$ & $X^+_5$ & $X^+_5$ & $X^+_5$ & $X^-_5$ & $X^-_5$ & $X^-_5$ & 
$X^-_5$\\
\hline\\
\end{tabular}\hspace{.7cm}
\begin{tabular}[t]{ccccc}
\multicolumn{5}{c}{  $P$}\\
\hline
$P_1$ & $P_2$ & $P_3$ & $P_4$ & $P_5$\\
$P_6$ & $P_6$ & $P_7$ & $P_7$ & $P_6$ + $P_7$\\
\hline\\
\end{tabular}
\end{table*}

\begin{table*}[H]
\caption{
  Compatibility relations between the single-valued (upper row) and
  double-valued (lower row) representations of the space group $Cmca$ of the
  orthorhombic phase of La$_2$CuO$_4$.  For the notations see Tables 
  \protect\ref{tab:edortho} and \protect\ref{tab:zdortho}. The
  small representations in the same 
  column are compatible in the following sense: spin-dependent Bloch
  functions which are basis functions of the
  small representation $R_i\times D_{1/2}$ (with $R_i$ standing in the upper
  row) also form basis functions of the double-valued representation
  standing 
  below $R_i$.  $D_{1/2}$ denotes the two-dimensional double-valued
  representation of the three-dimensional rotation group $O(3)$ (given,
  e.g., in Table 6.1 of Ref.~\protect\cite{bc}).
\label{tab:edzdortho}
}
\begin{tabular}[t]{cccccccc}
\multicolumn{8}{c}{  $\Gamma$}\\
\hline
$\Gamma^+_1$ & $\Gamma^+_2$ & $\Gamma^+_3$ & $\Gamma^+_4$ & $\Gamma^-_1$ & 
$\Gamma^-_2$ & $\Gamma^-_3$ & $\Gamma^-_4$\\
$\Gamma^+_5$ & $\Gamma^+_5$ & $\Gamma^+_5$ & $\Gamma^+_5$ & $\Gamma^-_5$ & 
$\Gamma^-_5$ & $\Gamma^-_5$ & $\Gamma^-_5$\\
\hline\\
\end{tabular}\hspace{.7cm}
\begin{tabular}[t]{cccccccc}
\multicolumn{8}{c}{  $Y$}\\
\hline
$Y^+_1$ & $Y^+_2$ & $Y^+_3$ & $Y^+_4$ & $Y^-_1$ & $Y^-_2$ & $Y^-_3$ & 
$Y^-_4$\\
$Y^+_5$ & $Y^+_5$ & $Y^+_5$ & $Y^+_5$ & $Y^-_5$ & $Y^-_5$ & $Y^-_5$ & 
$Y^-_5$\\
\hline\\
\end{tabular}\\
\begin{tabular}[t]{cc}
\multicolumn{2}{c}{  $Z$}\\
\hline
$Z_1$ & $Z_2$\\
$Z_3$ + $Z_4$ & $Z_3$ + $Z_4$\\
\hline\\
\end{tabular}\hspace{.7cm}
\begin{tabular}[t]{cc}
\multicolumn{2}{c}{  $T$}\\
\hline
$T_1$ & $T_2$\\
$T_3$ + $T_4$ & $T_3$ + $T_4$\\
\hline\\
\end{tabular}\hspace{.7cm}
\begin{tabular}[t]{c}
\multicolumn{1}{c}{  $S$}\\
\hline
$S_1$\\
2$S_2$\\
\hline\\
\end{tabular}\hspace{.7cm}
\begin{tabular}[t]{cccc}
\multicolumn{4}{c}{  $R$}\\
\hline
$R^+_1$ & $R^+_2$ & $R^-_1$ & $R^-_2$\\
$R^+_3$ + $R^+_4$ & $R^+_3$ + $R^+_4$ & $R^-_3$ + $R^-_4$ & $R^-_3$ + $R^-_4$\\
\hline\\
\end{tabular}
\end{table*}

\begin{table*}
\caption{
  Single-valued small representations (in the notation given in Table 
  \protect\ref{tab:edtetra}) of all the single bands in the space group
  $I4/mmm$ of 
  tetragonal La$_2$CuO$_4$ with symmetry-adapted and optimally localizable
  Wannier functions centered at the Cu atoms.
\label{tab:wftetra}
}
\begin{tabular}[t]{cp{.5cm}ccccc}
\hline
Band 1&& $\Gamma^+_1$ & $Z^+_1$ & 
$N^+_1$ & $X^+_1$ & $P_1$\\
Band 2&& $\Gamma^+_2$ & $Z^+_2$ & 
$N^+_2$ & $X^+_3$ & $P_2$\\
Band 3&& $\Gamma^+_3$ & $Z^+_3$ & 
$N^+_1$ & $X^+_3$ & $P_3$\\
Band 4&& $\Gamma^+_4$ & $Z^+_4$ & 
$N^+_2$ & $X^+_1$ & $P_4$\\
Band 5&& $\Gamma^-_1$ & $Z^-_1$ & 
$N^-_1$ & $X^-_1$ & $P_3$\\
Band 6&& $\Gamma^-_2$ & $Z^-_2$ & 
$N^-_2$ & $X^-_3$ & $P_4$\\
Band 7&& $\Gamma^-_3$ & $Z^-_3$ & 
$N^-_1$ & $X^-_3$ & $P_1$\\
Band 8&& $\Gamma^-_4$ & $Z^-_4$ & 
$N^-_2$ & $X^-_1$ & $P_2$\\
\hline\\
\end{tabular}
\end{table*}

\begin{table*}
\caption{
  (Double-valued) small representations (in the notation given in Table 
  \protect\ref{tab:zdtetra}) of all the superconducting bands in the space
  group 
  $I4/mmm$ of tetragonal La$_2$CuO$_4$ with symmetry-adapted and optimally
  localizable spin-dependent Wannier functions centered at the Cu atoms.
\label{tab:slbandtetra}
}
\begin{tabular}[t]{cp{.5cm}ccccc}
\hline
Band 1&& $\Gamma^+_6$ & $Z^+_6$ & 
$N^+_3$ + $N^+_4$ & $X^+_5$ & $P_6$ \\
Band 2&& $\Gamma^+_7$ & $Z^+_7$ & 
$N^+_3$ + $N^+_4$ & $X^+_5$ & $P_7$ \\
Band 3&& $\Gamma^-_6$ & $Z^-_6$ & 
$N^-_3$ + $N^-_4$ & $X^-_5$ & $P_7$ \\
Band 4&& $\Gamma^-_7$ & $Z^-_7$ & 
$N^-_3$ + $N^-_4$ & $X^-_5$ & $P_6$ \\
\hline
\end{tabular}
\end{table*}

\begin{table*}
\caption{
  (Double-valued) small representations (in the notation given in Table 
  \protect\ref{tab:zdortho}) of all the superconducting bands in the space
  group
  $Cmca$ of the orthorhombic phase of La$_2$CuO$_4$ with symmetry-adapted
  and optimally localizable spin-dependent Wannier functions centered at the
  Cu atoms. 
\label{tab:slbandortho}
}
\begin{tabular}[t]{cp{.5cm}cccccc}
\hline
Band 1&& 2$\Gamma^+_5$ & 2$Y^+_5$ & 
$Z_3$ + $Z_4$ & $T_3$ + $T_4$ & 2$S_2$ & 2$R^+_3$ + 2$R^+_4$\\
Band 2&& 2$\Gamma^-_5$ & 2$Y^-_5$ & 
$Z_3$ + $Z_4$ & $T_3$ + $T_4$ & 2$S_2$ & 2$R^-_3$ + 2$R^-_4$\\
\hline
\end{tabular}
\end{table*}

\begin{table*}[H]
\caption{
  (Single-valued) small representations (in the notation given in Table 
  \protect\ref{tab:edortho}) of all the neutral bands in the space
  group 
  $Cmca$ of the orthorhombic phase of La$_2$CuO$_4$ with symmetry-adapted
  and optimally localizable Wannier functions centered at the Cu atoms.
  The Wannier functions may be constructed so that either they are
  symmetry-adapted to the  
  magnetic group $M_1 = Cmca + \{K|\frac{1}{2}\frac{1}{2}\frac{1}{2}\}Cmca$
  or to the 
  grey magnetic group $M_2 = Cmca + K\!\cdot\! Cmca$ because Eq.~\gl{eq:1}
  is satisfied in both cases.  
  However, an antiferromagnetic structure with the magnetic group $M_1$ is
  unstable, see Appendix~\protect\ref{sec:condition}. Hence, the listed
  bands are neutral bands with Wannier functions adapted to $M_2$.
\label{tab:afbandortho}
}
\begin{tabular}[t]{cp{.5cm}cccccc}
\hline
Band 1&& $\Gamma^+_1$ + $\Gamma^+_4$ & 
$Y^+_1$ + $Y^+_4$ & $Z_1$ & $T_1$ & $S_1$ & $R^+_1$ + $R^+_2$ \\
Band 2&& $\Gamma^+_2$ + $\Gamma^+_3$ & 
$Y^+_2$ + $Y^+_3$ & $Z_2$ & $T_2$ & $S_1$ & $R^+_1$ + $R^+_2$ \\
Band 3&& $\Gamma^-_1$ + $\Gamma^-_4$ & 
$Y^-_1$ + $Y^-_4$ & $Z_2$ & $T_2$ & $S_1$ & $R^-_1$ + $R^-_2$ \\
Band 4&& $\Gamma^-_2$ + $\Gamma^-_3$ & 
$Y^-_2$ + $Y^-_3$ & $Z_1$ & $T_1$ & $S_1$ & $R^-_1$ + $R^-_2$ \\
\hline
\end{tabular}
\end{table*}

\begin{table*}[H]
\caption{
  (Single-valued) small representations (in the notation of
  Ref.~\protect\cite{eabf}) of all  
  the antiferromagnetic bands in the space group $P4/mnc = D^6_{4h}$ (128)
  of antiferromagnetic Cr with symmetry-adapted and optimally localizable
  Wannier functions centered at the Cu atoms. This table has already been
  published in Ref.~\protect\cite{ew4}.  
\label{tab:afbandcr}
}
\begin{tabular}[t]{cp{.5cm}cccccc}
\hline
Band 1&& $\Gamma^+_1$ + $\Gamma^+_2$ & $X_{5}$ & $M_{20}$ & 
$A_{11}$ & $Z_{10}$ & $R_{5}$\\
Band 2&& $\Gamma^+_3$ + $\Gamma^+_4$ & $X_{5}$ & $M_{20}$ & 
$A_{10}$ & $Z_{11}$ & $R_{5}$\\
Band 3&& $\Gamma^-_1$ + $\Gamma^-_2$ & $X_{10}$ & $M_{10}$ & 
$A_{10}$ & $Z_{10}$ & $R_{10}$\\
Band 4&& $\Gamma^-_3$ + $\Gamma^-_4$ & $X_{10}$ & $M_{10}$ & 
$A_{11}$ & $Z_{11}$ & $R_{10}$\\
\hline
\end{tabular}
\end{table*}

\begin{table*}
\caption{
  (Single-valued) small representations (in the notation given in Table 
  \protect\ref{tab:edaf}) of all the antiferromagnetic bands in the space
  group
  $Pccn$ of the (experimentally established) antiferromagnetic structure of
  La$_2$CuO$_4$ with symmetry-adapted and optimally localizable Wannier
  functions centered at the Cu atoms. An antiferromagnetic structure with
  the magnetic group $M = Pccn +
  \{K|\textstyle\frac{1}{2}\frac{1}{2}0\}Pccn$ 
  can be stable, see 
  Appendix~\protect\ref{sec:condition}. 
\label{tab:afband}
}
\begin{tabular}[t]{cp{.5cm}cccccccc}
\hline
Band 1&& $\Gamma^+_1$ + $\Gamma^+_2$
 + $\Gamma^+_3$ + $\Gamma^+_4$ & $Y_1$ + $Y_2$ & $X_1$ + $X_2$ & $Z_1$ + $Z_2$ & 
$U_1$ + $U_2$ & $T_1$ + $T_2$ & $S_1$ + $S_2$ & $R^+_1$ + $R^+_2$ + $R^+_3$
 + $R^+_4$\\
Band 2&& $\Gamma^-_1$ + $\Gamma^-_2$
 + $\Gamma^-_3$ + $\Gamma^-_4$ & $Y_1$ + $Y_2$ & $X_1$ + $X_2$ & $Z_1$ + $Z_2$ & 
$U_1$ + $U_2$ & $T_1$ + $T_2$ & $S_1$ + $S_2$ & $R^-_1$ + $R^-_2$ + $R^-_3$
 + $R^-_4$\\
\hline
\end{tabular}
\end{table*}

\end{document}